\documentclass[journal]{IEEEtran}
\usepackage[utf8]{inputenc}
\usepackage{bm}
\usepackage{amsfonts,amsmath,amssymb}
\usepackage{graphicx}
\usepackage{soul}
\usepackage{siunitx}
\usepackage{cite}
\usepackage{balance}
\usepackage{caption}
\usepackage{subfig} 
\usepackage{comment}
\usepackage{enumitem}
\usepackage{bbold}
\usepackage{booktabs}
\usepackage{multirow}
\usepackage[table,xcdraw]{xcolor}

\usepackage{pgfplots}
\pgfplotsset{compat=newest}
\usepgfplotslibrary{statistics}
\usepackage{tikz}
\usepackage{pgfplotstable}
\usetikzlibrary{shapes, arrows, positioning, backgrounds, calc, matrix}

\usepackage[acronym,shortcuts]{glossaries}
\newacronym{adam}{ADAM}{adaptive moment estimation}
\newacronym{aoa}{AOA}{angle-of-arrival}
\newacronym{rnn}{RNN}{recurrent neural network}
\newacronym{mfp}{MFP}{matched field processing}
\newacronym{nn}{NN}{neural network}
\newacronym{ffnn}{FFNN}{feed forward neural network}
\newacronym{svm}{SVM}{support vector machine}
\newacronym{grnn}{GRNN}{generalized regression neural network}
\newacronym{cnn}{CNN}{convolutional neural network}
\newacronym{scm}{SCM}{sample covariance matrix}
\newacronym{lstm}{LSTM}{long short term memory}
\newacronym{rmse}{RMSE}{root mean square error}
\newacronym{uwan}{UWAN}{underwater acoustic network}
\newacronym{uwac}{UWAC}{underwater acoustic channel}
\newacronym{uwa}{UWA}{underwater acoustic}
\newacronym{mape}{MAPE}{mean absolute percentage error}
\newacronym{mae}{MAE}{mean absolute error}
\newacronym{mse}{MSE}{mean squared error}
\newacronym{det}{DET}{detection error tradeoff}
\newacronym{wma}{WMA}{weighted moving average}
\newacronym{ml}{ML}{machine learning}
\newacronym{auv}{AUV}{autonomous underwater vehicle}
\newacronym{snr}{SNR}{signal to noise ratio}
\newacronym{ssp}{SSP}{sound speed profile}
\newacronym{glrt}{GLRT}{generalized likelihood ratio test}
\newacronym{std}{STD}{standard deviation}

\newlength\fwidth
\newlength\fheight
\makeatletter
\newcommand{\new}[1]{{\textcolor{blue}{#1}}}

\makeatletter
\newcommand\remembertext[2]{
  \immediate\write\@auxout{\unexpanded{\global\long\@namedef{mytext@#1}{#2}}}%
  {\color{blue} #2}%
}
\newcommand\recalltext[1]{%
  \new{\ifcsname mytext@#1\endcsname
    \fontsize{10.5}{12.5}\selectfont\@nameuse{mytext@#1}%
  \else
    ``??''
  \fi
}}
\makeatother

\title{Authentication by Location Tracking \\ in Underwater Acoustic Networks}
\author{Gianmaria Ventura, \IEEEmembership{Student Member, IEEE}, Francesco Ardizzon, \IEEEmembership{Member, IEEE}, \\ and Stefano Tomasin, \IEEEmembership{Senior Member, IEEE}\thanks{Manuscript received --; accepted --. Date --; date of current version --. Corresponding author: G. Ventura. This work was sponsored in part by the NATO Science for Peace and Security Programme under grant no. G5884 (SAFE-UComm).This work was also partially supported by the European Union under the Italian National Recovery and Resilience Plan (PNRR) of NextGenerationEU, partnership on “Telecommunications of the Future” (PE0000001 - program “RESTART”). The authors are with the Department of Information Engineering, Universit\`a degli Studi di Padova, Padua 35131, Italy. S. Tomasin is also with the National Inter-University Consortium for Telecommunications (CNIT), 43124 Parma, Italy. (email:  gianmaria.ventura@phd.unipd.it, francesco.ardizzon@phd.unipd.it, stefano.tomasin@unipd.it).}}

\begin{document}

\maketitle
\begin{abstract}
    Physical layer message authentication in \acp{uwan} leverages the characteristics of the \ac{uwac} as a fingerprint of the transmitting device. However, as the device moves its \ac{uwac} changes, and the authentication mechanism must track such variations. In this paper, we propose a context-based authentication mechanism operating in two steps: first, we estimate the position of the underwater device, then we predict its future position based on the previously estimated ones. To check the authenticity of the transmission, we compare the estimated and the predicted position. The location is estimated using a \acl{cnn} taking as input the \acl{scm} of the estimated \acp{uwac}. The prediction uses either a Kalman filter or a \ac{rnn}. The authentication check is performed on the squared error between the predicted and estimated positions. The solution based on the Kalman filter outperforms that built on the \ac{rnn} when the device moves according to a correlated Gauss-Markov mobility model, which reproduces a typical underwater motion. 
\end{abstract}


\glsresetall

\sloppy

\section{Introduction}\label{sec:intro}

\IEEEPARstart{R}{ecently}, interest in \glspl{uwan} has grown due to their application in contexts such as natural resources extraction, ocean exploration, military operations, climate change monitoring, and marine pollution control. However, as \gls{uwan} applications become more popular, security concerns also increase. 

\Ac{uwan} attacks can range from simple signal jamming to impersonation attacks~\cite{lal2016secure, yang2019challenges}.
The first solution to prevent potential attacks is to employ security protocols based on cryptography. These however have a higher communication overhead, further reducing the already low rate of data transmissions in \ac{uwan}. In this context, physical layer-based solutions that leverage the \gls{uwac} characteristics provide an interesting alternative, due to their minimal overhead and utilization of \ac{uwac} features already needed for communication purposes. The current authentication strategies for underwater acoustic communication have been surveyed in~\cite{Waquas23Security}. 

This paper focuses on authenticating messages exchanged between underwater devices with mechanisms operating at the physical layer.
Physical-layer authentication techniques are trust models based on ad-hoc chosen metrics, acting as fingerprints. The underlying assumption is that the statistical distribution of the security metrics depends on transmitter and receiver positions. Thus, the source can be authenticated by verifying the match between the metrics and their expected statistic.  Moreover, in this scenario, it is unrealistic to assume any knowledge of the attacker challenge distribution, as it is directly linked to the attacker's movement, which typically remains unknown.

In~\cite{diamant2018cooperative}, the authors proposed a set of metrics that are stable over time but not over space, thus a small shift in the transmitter/receiver's position causes relevant metric variations. Next, the metrics' distributions are estimated using a generalized Gaussian model. The model is later used to verify the channel authenticity by a \acrlong{glrt}. Additionally, the authors propose relying on node collaboration to improve performance.

Other metrics have been proposed in ~\cite{khalid_auth_aoa_mahala_oceans_2020, zhao22physical, Zhao2023Physical}. In particular, the measured angle of arrival has been used in~\cite{khalid_auth_aoa_mahala_oceans_2020}, the so-called  time-reversal resonating strength in~\cite{zhao22physical}, while the maximum and the minimum correlation with of the measured channel impulse response with respect to a previously collected (positive only) dataset is studied in~\cite{Zhao2023Physical}.
Concerning the method to verify the authenticity of the measurements, a common trend is to avoid the complex metrics' distribution estimation in favor of data-driven \gls{ml}-based checks~\cite{bragagnolo2021authentication,du2022ltrust,zhang2023recommendation}.
\IEEEpubidadjcol
 Authentication of a moving transmitter poses further challenges, as both channel and metric distribution change over time. A typical solution is then to frame this problem as anomaly detection, where the channel is continuously monitored, and unexpected behaviors and discontinuities are associated with the start of an attack.

Along this line of research, the authors proposed in~\cite{casari2022physical} a Kalman-based predictor to estimate the source position and velocity by tracking the power-weighted average of the channel taps delay, and the innovation of the Kalman filter is used to perform source authentication. The work was further extended in \cite{Ardizzon2024RNN}, where the Kalman filter is replaced by a \ac{rnn} to track the evolution of harder-to-estimate features. A similar approach is also proposed in \cite{Aman2024novel}, where the security metric is the time-of-arrival tracked by a Kalman filter. 


In this paper, instead, we propose a context-based authentication mechanism. This process involves explicitly determining the source position by tracking the device's movement and utilizing location as contextual information. Indeed, the device positions and thus its relative distance is valuable information that can also be exploited for other purposes beyond authentication, e.g., at the MAC layer for data transmission scheduling or for signal beamforming.

Thus, concerning localization in the underwater acoustic context, early works utilize the \gls{mfp} technique~\cite{sazontov2015matched} to compare the measured acoustic pressure field with a local replica. This technique relies on the knowledge of the signal propagation model that, in turn, may be dependent on environmental factors and on the source position to be estimated. However, obtaining this information may be challenging in many practical scenarios. 

\Gls{ml} models have emerged as recent solutions since source localization can be seen as a regression problem. These methods are data-driven and eliminate the need for environmental information or propagation models. The dataset can be collected from previously conducted experiments or simulators, such as Bellhop ray tracer~\cite{bellhop}. 
In~\cite{niu2017source}, \gls{nn}, \gls{svm}, and random forest techniques were exploited to predict the source range by taking as input the normalized \gls{scm} of complex acoustic pressure in the frequency domain. Indeed, such approaches often outperform the \gls{mfp}-based solution. Other localization methods are based on \gls{grnn}~\cite{wang2018underwater}, \gls{cnn}~\cite{huang2018source, liu2020source, liu2020multi}, and \gls{rnn} for both the underwater and more general communication contexts~\cite{9430737,qin2020underwater,zhu2022time}. 

In this paper, we introduce a novel source position prediction method exploiting an array of hydrophones. Such position estimation is then exploited for authentication purposes, following a context-based authentication approach. 
For localization, a \gls{cnn} is used to localize the source from the receivers' measurements. The \gls{scm} is selected as input to the \ac{cnn} as it compactly represents the spatial and frequency correlation between the receivers. Concerning the position tracker, two approaches have been considered, a Kalman filter and a \gls{rnn}-based predictor, trained to predict the current position based on past ones. For authentication, first, we employ a predictor to estimate the current location from the past ones. Then we compute the squared error between the predicted and observed positions. The selected metric is compared to a suitably chosen threshold to check authentication. 

More in detail, the paper's main contributions are as follows.
\begin{enumerate}
    \item We propose a lightweight \ac{cnn} that, rather than only measuring its distance, localizes the transmitter by using the covariance matrix of the channel frequency response measured by each sensor. 
    \item We design novel predictors to track the transmitter trajectory, namely a Kalman filter-based and an \ac{rnn}-based predictor. The two predictors show similar performances. 
    \item We exploit the predicted source position to distinguish between an authentic transmitter and an attacker.
    \item We test the proposed solution using a dataset generated via Bellhop, also modeling complex environmental characteristics such as water pH and temperature. 
\end{enumerate}

The rest of the paper is organized as follows. Section~\ref{sec:systModel} describes the system model. Section~\ref{sec:propProtocol} presents the proposed authentication protocol. Section~\ref{sec:numResults} details the simulation's setup and the localization and authentication results. Finally, Section~\ref{sec:concl} draws the conclusions.

\section{System Model}\label{sec:systModel}
We consider an \gls{uwan} composed of one transmitter e.g., a drifter or an autonomous underwater vehicle, Alice, and a set of $N_\mathrm{rx}$ static receivers, named Bob. All receivers are synchronized and connected to a {\em sink node} via an error-free, authenticated, and integrity-protected channel, used only to share their observations. Next, we consider a malicious transmitter, Eve that moves in the same area.

We consider a scenario where Bob may receive messages from Alice or Eve. Eve is transmitting messages impersonating Alice, thus pretending that her messages are coming from Alice, e.g., by placing Alice's address in her message data. Bob aims to verify if the received messages have been transmitted by Alice. To perform authentication, Bob exploits the estimate of the transmission channel over which messages are received. Such a channel depends on the position of the transmitter thus operating as its fingerprint.

\begin{figure} 
    \centering
    \includegraphics[width=.8\linewidth]{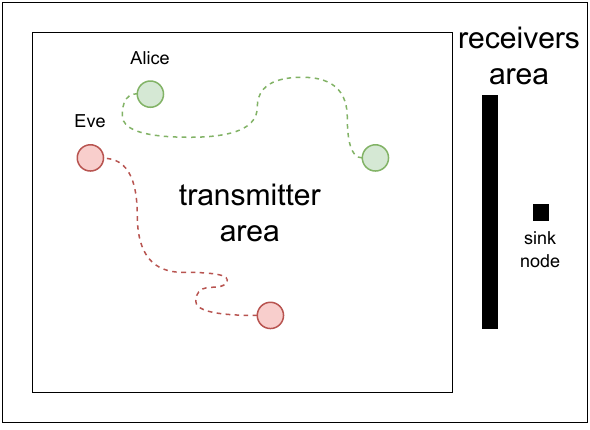}
    \caption{Example of Alice, Bob, and Eve's positions and movements.}
    \label{fig:example-disp}
    \vspace{-.5cm}
\end{figure}

For the sake of simplicity, we assume all the devices to be placed at depth $d$, thus a device position is uniquely identified by two Cartesian coordinates. Such a setting can be easily generalized to a three-dimensional (3D) problem, at the cost of a more involved notation. Specifically, we denote the source position, speed, and acceleration as the vectors 
$\bm{p}= (x \; y)$, $\bm{v}= (v_\mathrm{x} \; v_\mathrm{y})$, and $\bm{a}= (a_\mathrm{x} \; a_\mathrm{y})$, respectively. Fig.~\ref{fig:example-disp} shows an example of device positions and movements.
\begin{figure*}
    \centering
   
    \begin{tikzpicture}[auto, node distance=3cm,>=latex']
    \tikzstyle{every node}=[font=\small]

    \tikzstyle{block} = [draw, fill=white, rectangle, 
        minimum height=3em, minimum width=6em]
        
    \tikzstyle{delay} = [draw, fill=white, rectangle, 
        minimum height=2em, minimum width=2em]
        
    \tikzstyle{sum} = [draw, fill=white, circle, minimum size=0.1em,inner sep=0pt, outer sep=0pt, node distance=1cm]
    
    \tikzstyle{input} = [coordinate]
    \tikzstyle{output} = [coordinate]
    \tikzstyle{pinstyle} = [pin edge={to-,thin,black}]

    \node [input, name=input] {};
    \node [block, right=1cm of input](estimator) {Pos. Estimator};
    \node [delay, right of=estimator](delay){$T$};
    \node [block, right of=delay](predictor){Pos. Predictor};
    \node[sum, right=2cm of predictor](sum){$-$};

    \node [delay, right=.5cm of sum](norm){$\|\cdot \|_2^2$};

    \node [block, right=1cm of norm](decision){Decision};
    \node [input, name=output, right=1cm of decision] {};

    \draw [draw,->] (input) -- node {${\bm{C}}(t)$} (estimator);
    \draw [draw,->] (estimator) -- node[name=pos] {$\tilde{\bm{p}}(t)$} (delay);
    \draw [draw,->] (delay) -- node {$\tilde{\bm{p}}(t-T)$} (predictor);
    \draw [draw,->] (predictor) -- node {$\hat{\bm{p}}(t-T)$} (sum);
    
    \draw [draw,->] (sum) -- node {} (norm);
    \draw [draw,->] (norm) -- node {$\mathcal{E}(t)$} (decision);

    \node [input, name=mid, below= 0.5cm of delay] {};
    \draw [-] (pos) |- node [name=y] {}(mid);
    \draw [->] (mid) -| node[pos=0.99] {} (sum);
    \draw [draw,->] (decision) -- node {$\hat{\mathcal{H}}(t)$} (output);



        
\end{tikzpicture}
    \caption{Workflow of the proposed source localization and authentication protocol.}
    \label{fig:workflow}
    \vspace{-.5cm}
\end{figure*}
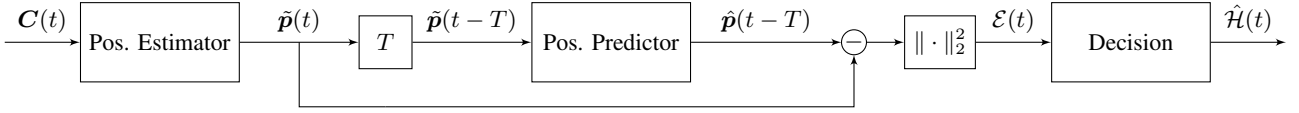

All devices communicate through a broadband \gls{uwac} with bandwidth $B$ centered at frequency $f_0$.
The transmission band is split into $K$ sub-bands and the channel is assumed to be narrowband within each sub-band. We denote with $H_{r,k}$ ($G_{r,k}$) the complex baseband equivalent narrowband channel between Alice (Eve) and receiver $r \in [1, N_\mathrm{rx}]$, for sub-band $k$ centered at frequency $f_k$. The noisy channel for receiver $r$ at sub-band $k$ is modeled as 
\begin{equation}
    Q_{r,k}= w_{r,k} + \left \{ 
    \begin{array}{rl}
    \begin{aligned}
        &H_{r,k}, \hspace{.5cm} \text{if Alice is transmitting,}
    \\
        &G_{r,k}, \hspace{.5cm} \text{if Eve is transmitting,}
    \end{aligned}
    \end{array} 
    \right.
\end{equation}

We collect the noisy channel gains of sub-band $k$ for every receiver into the column vector 
\begin{equation}    
\bm{Q}_k = (Q_{1,k}, \ldots, Q_{r,k}, \ldots, Q_{N_\mathrm{rx},k})^T,
\end{equation} 
and define its normalized counterpart as 
\begin{equation}
\bm{\Tilde{Q}}_k = \frac{\bm{Q}_k}{\|\bm{Q}_k\|_2},
\label{eq:norm}
\end{equation} 
where $\|\bm{Q}_k\|_2$ is the norm-2 of the vector.

The sink node computes the \gls{scm} at frequency $f_k$ as
\begin{equation}
    \bm{C}_k = \bm{\Tilde{Q}}_k \bm{\Tilde{Q}}_k^{\dag}\;,
    \label{eq:cov}
\end{equation}
where $^\dag$ is the Hermitian conjugate operator.
Finally, it arranges the \glspl{scm} in the 3D matrix \begin{equation}
    \bm{C} = (\bm{C}_1; \ldots;  \bm{C}_K),
\end{equation}
where semicolumns separate the matrices in the third dimension.We then build a dataset of $M$ consecutively sampled $\bm{C}$ when the source is moving. Such a dataset will be later used to train the \acp{nn}.



\subsection{Security Model}\label{sec:secModel}
Here, we detail the considered security model, which also includes the assumptions made on attacker Eve. We consider the worst-case scenario where, when Eve transmits, Bob receives messages only from her. For instance, such a condition can be obtained when Eve transmits at a higher power than Alice, or Eve waits for time slots when Alice is not transmitting.

We assume Eve to know both Alice's and Bob's positions. On the other hand, Eve must be far at least $D$ from Alice to not be detected by Alice, modeling for example the use of sonar by Alice. 
Still, Eve is capable of tracking and estimating Alice's movement. Thus, we will consider attacks where Eve tries to impersonate Alice by transmitting to Bob while mimicking Alice's motion pattern. 

At an initial stage, Alice can transmit $N_\mathrm{a}$ pilot signals that are authenticated by higher-layer mechanisms and are thus unpredictable. Such initial secure transmission is used to obtain the first set of authenticated features and then exploited to authenticate forthcoming transmission using only physical-layer features. Indeed, using a cryptographic authentication protocol adds communication overhead and computational complexity. 

During training, we assume that signals are protected by a higher-layer authentication mechanism, thus preventing adversarial (pollution) attacks. Moreover, we remark that such an attack would require the pre-compensation of all the $N_{\rm rx}$ channels from Eve to each receiver at the same time, which may be unfeasible in practice. On the other hand, even if Eve could imitate all the channels, the proposed mechanism would still be secure, since, for an effective attack, Eve should still perfectly predict and replicate Alice's movements.

Finally, since the communication between each receiver is authenticated, Eve can interfere only before the receivers' front end.

\section{Proposed Authentication Protocol}\label{sec:propProtocol}

The proposed authentication protocol relies on the samples' temporal correlation. To authenticate a transmission, Bob performs $M$  estimates of the \gls{uwac}. While, for the sake of simplicity, we assume that Bob performs its \gls{uwac} estimates at regular time intervals spaced by $T$, the proposed models could be modified to work with irregular time series. Thus, at each instant $t = \ell T, \ell = 1, \ldots, M$, a new packet is transmitted (by either Alice or Eve) and received by the set of hydrophones Bob. 

At an initial stage, Alice is transmitting  $N_\mathrm{a}$  packets to be used to obtain the first set of authenticated features and then exploited by the proposed authentication protocol. 


In particular, when receiving a new message at time $t>N_\mathrm{a}T$, Bob
\begin{enumerate}
        \item predicts the position  $\Hat{\bm{p}}(t)=\left(\Hat{x}(t) \; \Hat{y}(t)\right)$ from $\Tilde{\bm{p}}(t-T)$, using the solutions described in Sections~III.B and III.C;
        \item estimates $\bm{C}(t)$ from the \ac{uwac}, as detailed in Section~\ref{sec:systModel} and feeds it as input to the position estimator (described in Section~III.A), which in turn outputs  $\Tilde{\bm{p}}(t)=\left(\Tilde{x}(t) \; \Tilde{y}(t)\right)$;
    \item computes the squared error between the predicted and estimated positions as
    \begin{equation}
        \label{eq:auth_metric}
        \mathcal{E}(t) = \| \Hat{\bm{p}}(t) - \Tilde{\bm{p}}(t) \|_2^2; 
    \end{equation} 
    \item decides about the authenticity of the received message according to the following test
    \begin{equation}
        \hat{\mathcal H}(t)=\left \{ \begin{array}{rl}
    0, \hspace{.5cm} \mathrm{if} \hspace{.5cm} \mathcal{E}(t) < \lambda, \\
    1, \hspace{.5cm} \mathrm{if} \hspace{.5cm} \mathcal{E}(t) \geq \lambda,
    \end{array}
    \right.
    \end{equation} where $\hat{\mathcal H}(t) = 0$ means that the received message has been recognized as authentic (coming from Alice) and $\hat{\mathcal H}(t) = 1$ means that the received message has been recognized as fake (coming from Eve).
\end{enumerate} 
Fig.~\ref{fig:workflow} shows a block scheme of the proposed authentication protocol.

Called $\mathcal H(t) = 0$ the legitimate case and $\mathcal H(t) = 1$ the under-attack case, i.e., when Eve is transmitting, the overall aim of our authentication technique is to minimize the missed detection (MD) probability
\begin{equation}
\label{eq:pmd}
    p_\mathrm{MD}(t) = P\left(\hat{\mathcal H}(t)  = 0\big|   H(t) = 1\right)\,,
\end{equation}
for a fixed false alarm (FA) probability
\begin{equation}
\label{eq:pfa}
p_\mathrm{FA}(t) = P\left(\hat{\mathcal H}(t)  = 1\big|   H(t) = 1\right)\,.
\end{equation}

We remark that the computational cost of this solution is due to both training and inference. Only the former requires multiple function evaluations; however, it is performed once and may be run on a dedicated device, e.g., designed only for this task. The cost of each single evaluation is due only to the number of parameters of the \ac{nn}. Thus, to limit the impact of the proposed authentication mechanism computational cost, we will look at solutions with a limited number of parameters.

In the remaining part of this Section, we will detail the designed position estimator and the position predictor blocks.

\subsection{Position Estimation}\label{subsec:cnn}
The task of the position estimator is to match the \acp{scm} $\bm{C}$ to the source position. This can be modeled as a regression problem that we tackle with a \ac{cnn}.

\begin{figure}
    \centering
    \includegraphics[width=.8\linewidth]{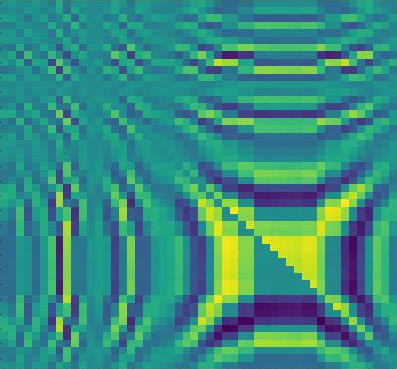}
    \caption{Example of processed \gls{scm} heatmap $\hat{\bm{C}}_k$.}
    \label{fig:covmatrix}
    \vspace{-.5cm}
\end{figure}

Before feeding the data to the \ac{cnn} we perform the following processing. First, we extract the upper triangular part of each matrix $\bm{C}_k(t)$ as $\bm{U}_k(t)$. Then, we build the full real-value matrix 
\begin{equation}
    \Hat{\bm{C}_k}(t) = \Re(\bm{U}_k(t)) + \Im(\bm{U}_k(t))^T,
\end{equation}
where $\Re(\cdot)$ and $\Im(\cdot)$ extract the real and imaginary parts of their argument, respectively. Next, we normalize the obtained matrix to have zero mean and unitary variance. An example of such a matrix is shown in Fig.~\ref{fig:covmatrix}.

The processed 3D matrix 
\begin{equation}
    \Hat{\bm{C}}(t) = \left( \Hat{\bm{C}}_1(t);...;\Hat{\bm{C}}_K(t)\right)
\end{equation}
is then used as input for the developed \ac{cnn}.

To solve the regression problem a \gls{cnn} called localization network (LOC-NET) is now introduced. The driving idea is that $K$ covariance matrices can be treated as an image having $K$ channels, instead of the 3 in RGB images. Indeed, a \gls{cnn} can exploit the spatial and channel (i.e., the frequency) correlation in image-like input data.
Inspired by the \gls{cnn} developed in~\cite{liu2020multi}, which is based on the Xception~\cite{Chollet_2017_CVPR} network for image classification, we introduce the LOC-NET architecture. We remark that LOC-NET is a novel architecture, as it inherits only the convolutional block structure from the other \ac{cnn}. Additionally, we do not use residual connections since LOC-NET is much smaller than the \gls{cnn} of~\cite{liu2020multi}. Furthermore, the smaller \ac{nn} yields faster and easier training and makes it suitable for \acp{uwan}. 

\paragraph*{LOC-NET Architecture} In the \ac{cnn} LOC-NET, each convolutional block is composed of: 
\begin{itemize}
    \item a 2D convolutional layer,
    \item a batch normalization layer,
    \item a ReLu activation function,
    \item a dropout layer.
\end{itemize}
Using convolutional layers instead of dense layers reduces the number of learnable parameters while the batch normalization layer avoids the internal covariate shift and makes the training faster and more stable~\cite{ioffe2015batch}. 
ReLu's are selected as the standard activation functions in \gls{cnn}~\cite{hao2020role} while the dropout layer prevents overfitting during the training procedure~\cite{srivastava2014dropout} and improves the network generalization abilities. The output of the \gls{cnn} $\Tilde{\bm{p}}(t)$ is an estimate of the transmitter position at time $t$, $\bm{p}(t)$. 

\paragraph*{LOC-NET Training} During the training, the \gls{cnn} is expected to learn the map between each source position and \glspl{scm}. To this end, we assume to have a database containing both $\bm{C}(t)$ and the corresponding true position $\bm{p}(t)$. For this regression problem, we consider as loss function the \ac{mse} 
\begin{equation}
    l = \frac{1}{b}\sum_{i=1}^b{(x_i-\Tilde{x}_i)^2} + \frac{1}{b}\sum_{i=1}^b{(y_i-\Tilde{y}_i)^2},
    \label{eq:loss}
\end{equation}
where $b$ is the batch size and the index $i$ represents the $i$-th sample of the selected batch.

In the following, we detail two alternative architectures for position prediction, one based on a Kalman filter and the other on a \ac{rnn}. They both aim at predicting the location $\hat{\bm{p}}(t)$ from the output of the LOC-NET $\tilde{\bm{p}}(t-T)$.

\subsection{Position Prediction by Kalman Filter}\label{subsec:kalman}
The Kalman filter provides a computational-efficient  (recursive) solution based on the least-squares method~\cite{welch1995introduction}.
The Kalman filter is unsupervised, and, differently from typical ML-based solutions, it does not need extensive training procedures. However, in general, it requires prior knowledge about the source motion and the observation model.

As described in Section~\ref{sec:systModel}, we assume that Alice is moving on a horizontal plane at depth $d$. Thus, the Kalman filter tracks a transmitter moving on a 2D plane at a constant speed. 

The vector of position measurements  $\Tilde{\bm{p}}(t)^T $ collects the noisy estimates of coordinates provided by the LOC-NET. We  define the hidden state vector collecting the position and speed components estimated by the Kalman filter as 
\begin{equation}
    \boldsymbol{h}(t) = \left( \Hat{x}(t) \; \Hat{v}_x(t) \; \Hat{y}(t) \; \Hat{v}_y(t) \right)^T.
\end{equation}
The state transition matrix $\bm{F}$ and the observation matrix $\bm{O}$ are
\begin{equation}    
\bm{F} = \begin{pmatrix} \bm{X} & \bm{0} \\ \bm{0} & \bm{Y}\end{pmatrix}\,, \mbox{with } \quad  \bm{X} = \bm{Y} =  
\begin{pmatrix}
    1 & T\\
    0 & 1
\end{pmatrix}
\end{equation}
and
\begin{equation}
\bm{O} = \begin{pmatrix}1 & 0 & 0 & 0 \\ 0 & 0 & 1 & 0\end{pmatrix}.
\end{equation}
We call the (set a priori) noise covariance matrices of process and measurement $\bm{Q}$ and $\bm{R}$, respectively.

Both $\bm{h}(t)$ and $\bm{P}(t)$ have to be initialized, we set $\bm{h}(0) = \left( \Tilde{x}(0) \; 1 \; \Tilde{y}(0) \; 1\right)$ and its covariance $\bm{P}(0) = 10^3 \cdot \mathbb{1}_4$. At each considered time step $t= \ell T$ the following operations are performed.
First, during the \emph{time update}, the current state vector and the covariance matrix are predicted from the previous ones, as 
    \begin{equation}
    \label{eq:kalman}
        \bm{h}(t) = \bm{F} \bm{h}(t-T),
    \end{equation}
    \begin{equation}
        \bm{P}(t) = \bm{F} \bm{P}(t-T) \bm{F}^T + \bm{Q}.
    \end{equation}

When a new measurement is available, during the \emph{measurement update} step we correct the predicted state vector and the covariance matrix by using the new measurement $\Tilde{\bm{p}}(t)^T$, as 
\begin{equation}
        \bm{K} = \bm{P}(t) \bm{O}^T \left(\bm{O} \bm{P}(t) \bm{O}^T + \bm{R}\right)^{-1},
    \end{equation}
    \begin{equation}
    \label{eq:kalman2}
        \bm{h}(t) = \bm{h}(t) + \bm{K} \left(\Tilde{\bm{p}}(t)^T - \bm{O} \bm{h}(t)\right),
    \end{equation}
    \begin{equation}
        \bm{P}(t) = \left(\mathbb{1}_4 - \bm{K} \bm{O}\right) \bm{P}(t) \left(\mathbb{1}_4 - \bm{K} \bm{O}\right)^T + \bm{K} \bm{R} \bm{K}^T,
    \end{equation}
where $\bm{K}$ is called Kalman gain, which determines the amount of correction to be applied to the prediction.

Therefore, $\forall t>0$ the predicted position can be extracted from \eqref{eq:kalman} and \eqref{eq:kalman2} as
\begin{equation}
    \Hat{\bm{p}}(t) = \bm{O} \bm{F} \left[\bm{h}(t-T) + \bm{K} \left(\Tilde{\bm{p}}(t-T)^T - \bm{O} \bm{h}(t-T)\right)\right],
\end{equation}
where $\Tilde{\bm{p}}(t-T)^T$ is the source position estimate performed by the LOC-NET at time $t-T$.

The described Kalman filter works with regular time series of data, sampled with period $T$; however, it could be adapted to work also with irregular time series of data assuming that the time intervals between one sample and the next one are known. More details concerning the Kalman filter and its statistical properties can be found in~\cite[Ch. 13]{Kay:1993}.

\subsection{Position Prediction by Recurrent Neural Network}\label{subsec:rnn}
To circumvent the need for the a priori knowledge of the transmitter movement's model, we consider a data-driven model based on a \ac{rnn}, trained to predict the next source position $\hat{\bm{p}}(t)$ from the previous ones $\Tilde{\bm{p}}(t-T)$. In this way, the model is comparable with the Kalman and improves its performance by exploiting its inner memory.

In particular, we consider a \ac{rnn} comprising $N_\mathrm{\ell}$ \gls{lstm} layers that, with their internal memory, extract the sequences' temporal correlation, a dropout layer to avoid overfitting, and $N_\mathrm{d}$ dense layers. 

Each \gls{lstm} layer~\cite{sak2014long} is characterized by a {\em state} which is updated at each new input, with period $T$. In particular, input sample $\tilde{\bm{p}}(t-T)$ is processed using state at time $t-T$, which include a {\em cell state} $\bm{c}_{t-T}$ and a {\em hidden state} $\bm{h}_{t-T}$. The corresponding output of the layer is $\hat{\bm{p}}(t)$. Internally, first, the {\em input gate $\bm{i}_t$, the forget gate $\bm{f}_t$, and the cell gate $\bm{g}_t$} vectors are computed as follows. Denoting the sigmoid and hyperbolic activation functions as $\sigma_{\rm act}$ and $\tanh$ and the Hadamard product operator as $\odot$, we have  
\begin{IEEEeqnarray}{cl}
    \bm{i}_t &= \sigma_{\rm act}(\bm{W}_{ii}\tilde{\bm{p}}(t-T)+\bm{b}_{ii}+\bm{W}_{hi}\bm{h}_{t-T}+\bm{b}_{hi})\:,\\
    \bm{f}_t &= \sigma_{\rm act}(\bm{W}_{if}\tilde{\bm{p}}(t-T)+\bm{b}_{if}+\bm{W}_{hf}\bm{h}_{t-T}+\bm{b}_{hf})\:,\\
    \bm{g}_t &= \tanh(\bm{W}_{ig}\tilde{\bm{p}}(t-T)+\bm{b}_{ig}+\bm{W}_{hg}\bm{h}_{t-T}+\bm{b}_{hg})\:,
\end{IEEEeqnarray}
where $\bm{W}_{ii}$, $\bm{W}_{if}$, $\bm{W}_{ig}$, $\bm{W}_{hi}$, $\bm{W}_{hf}$, and $\bm{W}_{hg}$ are weight matrices, while $\bm{b}_{ii}$, $\bm{b}_{if}$, $\bm{b}_{ig}$, $\bm{b}_{hi}$, $\bm{b}_{hf}$, and $\bm{b}_{hg}$ are bias vectors. Then, the output of the layer is computed as follows
\begin{equation}
    \hat{\bm{p}}(t) = \sigma_{\rm act}(\bm{W}_{io}\tilde{\bm{p}}(t-T)+\bm{b}_{io}+\bm{W}_{ho}\bm{h}_{t-T}+\bm{b}_{ho})\:,
\end{equation}
where $\bm{W}_{io}$ and $\bm{W}_{ho}$ are weight matrices, while $\bm{b}_{io}$ and $\bm{b}_{ho}$ are bias vectors. 

Moreover, the cell and hidden states are updated for each input. In particular, the cell state is computed from the forget gate, which establishes how much information has to be kept from the previous cell state, the input, and the cell gate, as
\begin{equation}
    \label{eq:rnn}
    \bm{c}_t = \bm{f}_t\odot \bm{c}_{t-T}+\bm{i}_t\odot \bm{g}_t\:.
\end{equation}
Finally, the hidden state represents the output of the cell and is computed from both the cell state and the output gate to get direct information about the previous inputs, as 
\begin{equation}
\label{eq:rnn2}
    \bm{h}_t = \hat{\bm{p}}(t) \odot \tanh(\bm{c}_t)\:.
\end{equation} 

After proper training and $\forall t>0$, the proposed \ac{rnn} takes as input the source position estimated by the LOC-NET $\tilde{\bm{p}}(t-T)$, and it predicts the transmitter position $\hat{\bm{p}}(t)$ acting in the same way as the Kalman filter.


We remark that, even if we focus on the case where the data is sampled with a constant rate, \acp{rnn} can be adapted to irregular time series, as shown in~\cite{weerakody2021review}.

\section{Numerical Results}\label{sec:numResults}
In this Section, we present the performance of the proposed solution. First, we describe the dataset generation. Then, we detail the parameter design of the employed solution. Lastly, we evaluate the localization and authentication accuracy. Note that since all existing literature has only considered the authentication in a static scenario, we do compare our solution with existing approaches, which would perform very poorly when the source is moving.

\subsection{Dataset Generation: Bellhop Simulation}\label{subsec:bellhop}

The 3D underwater acoustic simulator Bellhop~\cite{bellhop}, a beam-tracing model for simulating acoustic pressure field propagation in ocean environments, was used to generate the datasets.

Two areas of the San Diego Bay have been selected for the simulation. To confine the transmitter movement, we selected an inner area where the transmitter was allowed to move. The receivers have been located just outside of the transmitter area (see Fig.~\ref{fig:example-disp}). 

During a simulation, the source travels inside the transmitter area at depth $d =
\SI{50}{\meter}$ according to a correlated Gauss-Markov mobility model. Specifically, it starts moving from a position selected uniformly at random in the area, with initial speed direction drawn uniformly random in $[0, 2\pi)$, and modulus $\bm{v}(0) = \SI{2}{\meter/\second}$. The source speed and position are then periodically updated with period $T= \SI{10}{\second}$ as follows 
\begin{equation}
    \label{eq:speed}
    \bm{v}(t+T) = \alpha \bm{v}(t) + \bm{\eta}(t) \sqrt{1-\alpha^2},
\end{equation}
\begin{equation}\label{eq:position}
    \bm{p}(t+T) = \bm{p}(t) + \bm{v}(t) T,
\end{equation}
where $\alpha = 1- 2\cdot 10^{-3}$ is the trajectory correlation factor, and $\bm{\eta}(t)$ is a 2D vector with zero-mean independent Gaussian entries and \acp{std}, respectively, $\sigma_x = \sigma_y = \SI{2}{\meter/\second}$. 

For each path, an acoustic signal with carrier frequency $f_0$ is transmitted from the source. The signal propagation is simulated through the 3D geometric beams approximation~\cite{porter1994finite}, which transmits beams for every couple of the selected elevation and bearing angles.
Since the metrics of interest are the complex amplitudes and the delays of the channel impulse response, Bellhop is run in the so-called {\em arrivals mode}.

\begin{figure}
        \centering
        \begin{tikzpicture}
    \begin{axis}[
    ylabel={Y coordinates [m]},
    xlabel={X coordinates [m]},
    ymin=0, ymax=2550,
    ytick distance=500,
    xmin=0, xmax=2400,
    xtick distance=500,
    xlabel style={font=\scriptsize}, ylabel style={font=\scriptsize}, ticklabel style={font=\scriptsize},
    axis line style = {white}
    ]
    
    \addplot [on background layer] graphics
       [xmin=0,ymin=0,xmax=2125,ymax=2550]
       {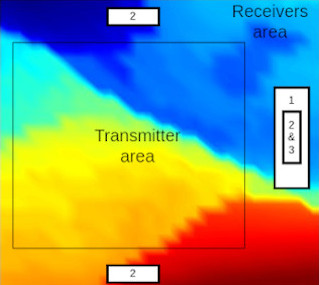};
    \end{axis}

    \begin{axis}[
        ymin=85, ymax=490,
        ytick distance=50,
        xmin=0,
        xmax=1,
        hide x axis,
        axis y line*=right,
        ylabel={\footnotesize{depth [m]}},
        ylabel shift = -3pt,
        ylabel style={font=\footnotesize},
        yticklabel style = {/pgf/number format/fixed, ticklabel style={font=\footnotesize}},
        axis line style = {white},
        every y tick/.style={black}
    ]
    \addplot graphics
       [xmin=0.92,ymin=85,xmax=1,ymax=490]
       {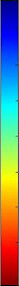};
  \end{axis}
\end{tikzpicture}
        \vspace{-.7cm}
        \caption{Receivers' dispositions 1, 2, and 3 for Bathymetry 1.}
        \vspace{-.3cm}
        \label{fig:dispositions}
\end{figure}

Fig.~\ref{fig:dispositions} shows the bathymetry used to produce the simulated dataset, where the corner with Cartesian coordinates $(0,0)$  has coordinates ($32^\circ52'30''$ N, $117^\circ19'25''$ W) with the $x$ and $y$ axis pointing respectively to the North and West. 
%
%
The size of the selected area is $2125~\mathrm{m}\times2550~\mathrm{m}$, with the transmitter placed in an inner area of dimensions $\SI{1500}{\meter}\times\SI{2000}{\meter}$ with corners' coordinates $(\SI{313}{\meter},\SI{275}{\meter} - (\SI{313}{\meter},\SI{2275}{\meter}) - (\SI{1813}{\meter},\SI{2275}{\meter}) - (\SI{1813}{\meter},\SI{275}{\meter})$. 
Furthermore, another area of the same size, called bathymetry 2, was selected $\SI{500}{\meter}$ North of the previous one to validate the proposed authentication model in a different geological setting.
In this setup, three different receivers' dispositions were tested:

\begin{description}
    \item[Disp. 1] $N_\mathrm{rx}=50$ receivers are uniformly placed along the line (\SI{1970}{\meter}, \SI{875}{\meter} -\SI{1375}{\meter});
    \item[Disp. 2] $N_\mathrm{rx}=50$ receivers are uniformly placed in a u-like shape. Ten of them are at coordinates (\SI{1970}{\meter}, \SI{1260}{\meter} - \SI{1320}{\meter}), twenty at (\SI{970}{\meter} - \SI{1160}{\meter}, \SI{2410}{\meter}) and the remaining at (\SI{970}{\meter} - \SI{1160}{\meter}, \SI{140}{\meter});
    \item[Disp. 3] $N_\mathrm{rx}=10$ receivers are uniformly placed along the line (\SI{1970}{\meter}, \SI{1230}{\meter} - \SI{1320}{\meter}).
\end{description}
In all dispositions, the receivers are placed at depth $d = \SI{50}{\meter}$. The dispositions have been sketched in Fig.~\ref{fig:dispositions}.
In addition, to evaluate the proposed model's robustness to the number of available receivers disposition $1$ has been modified with $N_{\rm rx} = 20,30$, and $40$.

Table~\ref{tab:param} summarizes the parameters to be set for the Bellhop simulator. More in detail,
\begin{itemize}
    \item the central frequency is $f_0 =\SI{11.5}{\kilo\hertz}$ with bandwidth $B = \SI{5}{\kilo\hertz}$, thus accommodating the $[9,14]\,$kHz band typically employed in \gls{uwa} communications;
    \item the noise variance for each receiver  at the desired \ac{snr} is computed as $\sigma_r^2 = {P_{r}}/{10^{\frac{\rm SNR}{10}}}\;,$
    where $P_{r}$ is the average received power for receiver $r$ at a distance between $\SI{200}{\meter}$ and $\SI{300}{\meter}$ from the transmitter;
    \item to achieve a trade-off between frequency accuracy and input data size we considered $K = 48$ frequencies;
    \item to ensure that the source beams reach the receivers from every position in the transmitter area we selected as elevation angles $\alpha_1 = -10^{\circ}$ and $\alpha_2 = 20^{\circ}$ and as bearing angles $\beta_1 = -90^{\circ}$ and $\beta_2 = 90^{\circ}$.
\end{itemize}
Each simulation is iterated for $M = 50$, hence the output of each simulation is a $50$ element cell composed by $N_\mathrm{rx}\times N_\mathrm{rx}\times K$ complex-valued matrices.

\begin{table}
    \centering
    \caption{Parameters for the Bellhop simulator.}
    \renewcommand{\arraystretch}{1.2} 
    \begin{tabular}{|c|c|}
       \hline
       \multicolumn{2}{|c|}{Bellhop parameters} \\
       \hline 
       carrier frequency  & $f_0\,$ \\
       sound speed profile  & Munk profile \\
       bottom boundary & acoustic-elastic half-space \\
       source coordinates & $p(t)$\\
       source/receiver depth & $d$\\
       receivers ranges & $N_\mathrm{rx}$ distances from source \\
       receivers bearings & $N_\mathrm{rx}$ bearing angles \\
       run type & arrivals, 3D, geometric beams \\
       elevation angles & 25 angles, from $\alpha_1$ to $\alpha_2$ \\ 
       bearing angles & 4 angles, from $\beta_1$ to $\beta_2$ \\ 
       bathymetry & San Diego Bay bathymetry (1 or 2) \\
       \hline
    \end{tabular}
    \label{tab:param}
\end{table}

\subsection{Model Architectures and Training Parameters}
\label{sec:architectures}
In this Section, we detail the design parameter of the LOC-NET, i.e., the position estimator, and both Kalman filter and \ac{rnn} which implement, in turn, the position predictor.

\paragraph*{LOC-NET}

\begin{figure}
    \centering
    \includegraphics[width=\linewidth]{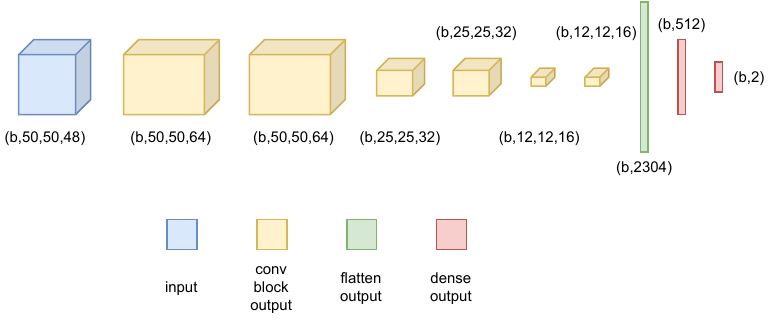}
    \caption{LOC-NET graphical representation.}
    \label{fig:conv_net}
    \vspace{-.5cm}
\end{figure}

After the preprocessing (see Section \ref{sec:propProtocol}), the input fed into the \gls{cnn} is a tensor of size $(b, N_\mathrm{rx}, N_\mathrm{rx}, K)$ where $K=48$, $b = 40$ is the selected batch size\footnote{The hyperparameter $b = 40$ was chosen via manual tuning.}, and $N_\mathrm{rx} =10 $ or $50$ is the number of receivers. It is worth noticing that two different LOC-NET architectures are proposed for $N_{\rm rx} =50$ and $N_{\rm rx} =10$. While for $N_{\rm rx} = 20,30$, and $40$, the input tensor is zero-padded to keep the same \ac{cnn} architecture as in the case with $N_{\rm rx} =50$. Moreover, since the LOC-NET learns how the transmitted signal propagates given an environment and a receiver's disposition, when the bathymetry or the receivers' disposition changes a new model has to be trained.
Then the input goes through six convolutional blocks composed of different layers, as described in Section~\ref{subsec:cnn}. 
After the convolutional blocks,  two dense layers provide as final output a $b \times 2$ matrix, collecting the predicted source position $\Hat{\bm{p}}$ for every sample in the considered batch. 
Indeed, the $x$ and $y$ coordinates are sufficient to identify the transmitter position as we assumed Alice can only move in the horizontal plane at depth $d = \SI{50}{\meter}$.
The architecture is sketched in Fig.~\ref{fig:conv_net}. 
We remark that the proposed \ac{cnn} has approximately \num{1.2e6} parameters while other \acp{cnn} performing source localization in simpler scenarios are much more complex. For example, different \ac{nn} models are proposed in~\cite{wang2018underwater}, with a number of parameters between \num{1.2e6} and \num{6.2e6}; the \ac{nn} developed in~\cite{huang2018source} has approximately \num{16e6} parameters, and the \ac{nn} in~\cite{liu2020multi} has even \num{5e9} parameters. 
Concerning the training, we used the \gls{mse} loss function~\eqref{eq:loss} and the \ac{adam} optimizer with weight decay parameter $10^{-5}$. 
The learning rate was initially set to $10^{-3}$ and then gradually decreased to $10^{-4}$ during the $300$ training epochs.
The best model was selected using a validation set, in this way, the generalization abilities of the model are preserved.
The training and validation datasets for the LOC-NET are respectively composed of $4000$ and $1000$ samples.

\paragraph*{RNN}
\begin{figure}
    \centering
    \includegraphics[width=0.8\linewidth]{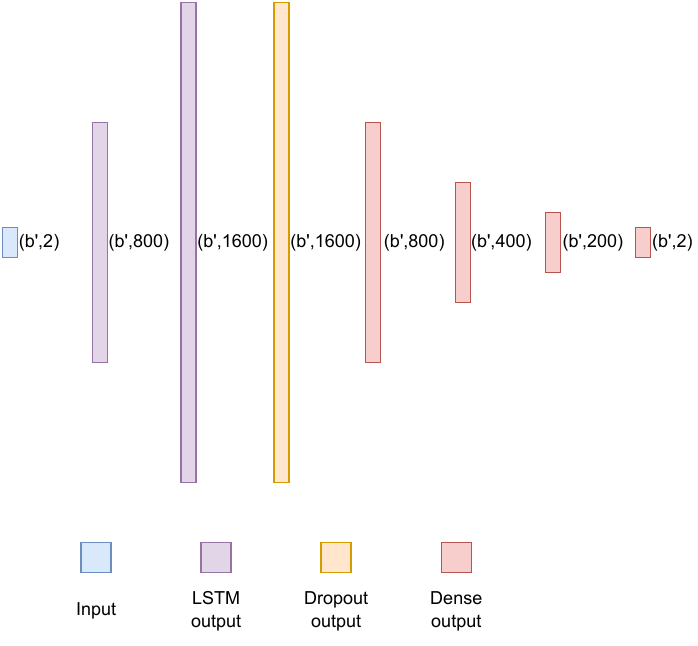}
    \caption{\ac{rnn} graphical representation.}
    \label{fig:rnn_net}
\end{figure}
The input of the \gls{rnn} is a tensor having shape $(b',2)$ with $b' = 8$ being the \ac{rnn}'s batch size optimum value.
The core features of the \gls{rnn} are two \gls{lstm} layers. The first has input size $2$ and hidden size $800$, while the second one has input size $800$ and hidden size $1600$. After the \ac{lstm} layer, we include a dropout layer. Then, after that, four dense layers gradually reduce the tensor shape to get an output of size $(b',2)$. This represents the source position at the next time step, for every sample in the considered batch.
Again, the \ac{mse} was selected as the loss function, \ac{adam} as optimizer with a weight decay parameter\footnote{Lower values led to model overfitting.} of $10^{-3}$. The learning rate of the optimizer starts again from  $10^{-3}$ but decreases to $10^{-5}$ along the $300$ training epochs.
Finally, a validation set was included to select the best model.
The training and validation datasets for the \ac{rnn}, instead, are respectively composed of the LOC-NET training and validation outputs.

\subsection{Localization Performance}
\label{sec:source-localization}
In this Section, we present the localization performance achieved by the different models described in Section~\ref{sec:propProtocol}. 

First, we investigate the localization performance of the LOC-NET, starting from the impact of the receivers' spatial distribution on localization. In particular, we consider the three dispositions of Section~\ref{subsec:bellhop} with ${\rm SNR} = \SI{20}{\decibel}$. Results are collected in Table~\ref{tab:disp_errors}.
As expected, dispositions 1 and 2 reach a higher accuracy than disposition 3, due to the higher number of receivers, i.e., $N_\mathrm{rx}=50$ vs. $N_\mathrm{rx}=10$. Still, when looking at applications with stricter constraints on $N_\mathrm{rx}$, it may be possible to seek an optimal tradeoff between the number of used receivers and the localization error. On the other hand, the similar average position errors achieved for dispositions 1 and 2 suggest that the geometric disposition of the receivers has a limited impact on the performance of the proposed estimator.  
\begin{table}
    \caption{Loc-net transmitter position error in different dispositions.}
    \centering
    \renewcommand{\arraystretch}{1.2} 
    \begin{tabular}{|c|c|c|c|}
        \hline
        Disposition N. & 1 & 2 & 3  \\
        \hline
        Avg. Localization error & $74$~m & $72$~m & $117$~m  \\

        \ac{std} Localization error & $58$~m & $50$~m & $114$~m \\
        \hline
    \end{tabular}
    \label{tab:disp_errors}
\end{table}

\begin{figure}
    \centering
    \subfloat[Disposition 1\label{fig:nn_heatmap_bathy1_disp1}]{%
        \begin{tikzpicture}
    \begin{axis}[
    ylabel={Y coordinates [m]},
    xlabel={X coordinates [m]},
    ymin=275, ymax=2275,
    ytick distance=200,
    xmin=313, xmax=1999,
    xtick distance=200,
    xlabel style={font=\scriptsize}, ylabel style={font=\scriptsize}, ticklabel style={font=\scriptsize},
    axis line style = {white}
    ]
    \addplot [on background layer] graphics
       [xmin=313,ymin=275,xmax=1813,ymax=2275]
       {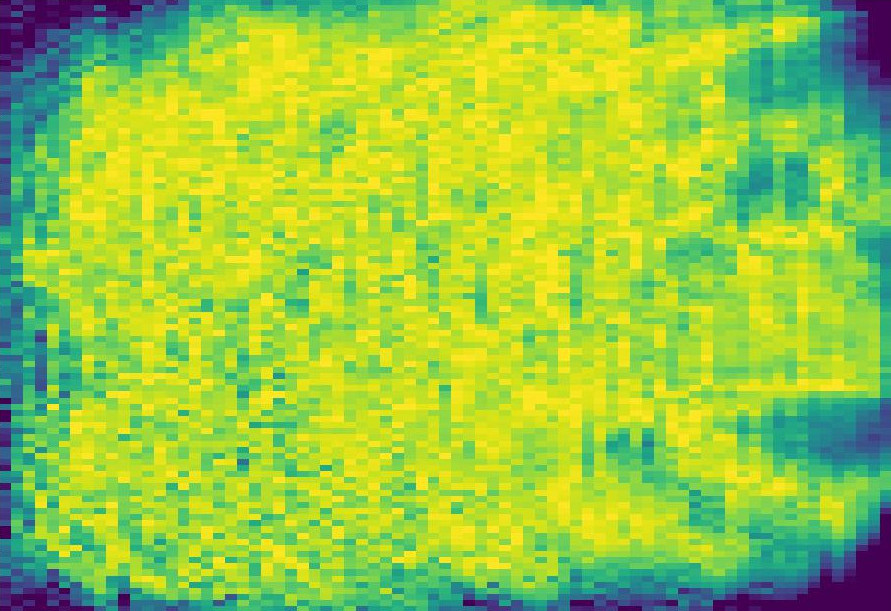};
    \end{axis}

    \begin{axis}[
        ymin=-1, ymax=451,
        ytick distance=50,
        xmin=0,
        xmax=1,
        hide x axis,
        axis y line*=right,
        ylabel={\footnotesize{RMSE [m]}},
        ylabel shift = -3pt,
        ylabel style={font=\footnotesize},
        yticklabel style = {/pgf/number format/fixed, ticklabel style={font=\footnotesize}},
        axis line style = {white}
    ]
    \addplot [on background layer] graphics
       [xmin=0.93,ymin=-1,xmax=1,ymax=451]
       {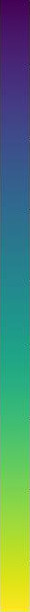};
  \end{axis}
\end{tikzpicture}
        }
    \\
    \subfloat[Disposition 2\label{fig:nn_heatmap_bathy1_disp2}]{%
        \begin{tikzpicture}
    \begin{axis}[
    ylabel={Y coordinates [m]},
    xlabel={X coordinates [m]},
    ymin=275, ymax=2275,
    ytick distance=200,
    xmin=313, xmax=1999,
    xtick distance=200,
    xlabel style={font=\scriptsize}, ylabel style={font=\scriptsize}, ticklabel style={font=\scriptsize},
    axis line style = {white}
    ]
    
    \addplot [on background layer] graphics
       [xmin=313,ymin=275,xmax=1813,ymax=2275]
       {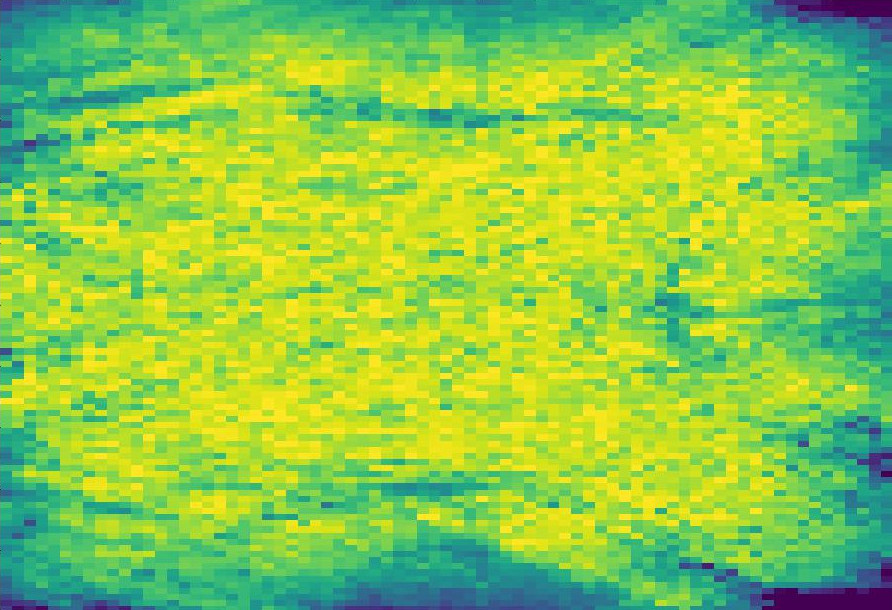};
    \end{axis}

    \begin{axis}[
        ymin=-1, ymax=451,
        ytick distance=50,
        xmin=0,
        xmax=1,
        hide x axis,
        axis y line*=right,
        ylabel={\footnotesize{RMSE [m]}},
        ylabel shift = -3pt,
        ylabel style={font=\footnotesize},
        yticklabel style = {/pgf/number format/fixed, ticklabel style={font=\footnotesize}},
        axis line style = {white}
    ]
    \addplot [on background layer] graphics
       [xmin=0.93,ymin=-1,xmax=1,ymax=451]
       {images/2heatmap_bathy1_disp1.jpeg};
  \end{axis}
\end{tikzpicture}
        }
    \caption{LOC-NET error heatmaps for disposition 1 and 2, for all the possible source positions.}
     \label{fig:nn_heatmap}
     \vspace{-.2cm}
\end{figure}

To further investigate the spatial distribution of errors, Fig.~\ref{fig:nn_heatmap} shows the localization errors of the LOC-NET, as a function of the transmitter position. In particular, the color of each point represents the magnitude of the LOC-NET localization error achieved when the source transmits from that point.
From Fig.~\ref{fig:nn_heatmap_bathy1_disp1} we observe that the transmitter positions leading to the larger errors are around the four corners, i.e., either locations almost aligned with the receiver line (right corners) or too far away from the receivers (left corners). Additionally, a wave-like behavior in the error is observed, parallel to the receivers' position. This confirms the idea that sources placed close along the y-axis will be hardly distinguished, due to the particular receiver placement.
Concerning disposition 2, reported in Fig.~\ref{fig:nn_heatmap_bathy1_disp2}, where the receivers are arranged in a {u-shape} around the transmitter area, while the highest errors are placed again at the corners, the error appears to be slightly more uniformly distributed. Still, in conclusion, we can state that higher errors occur when the transmitter is too far or too aligned with the receiver line.


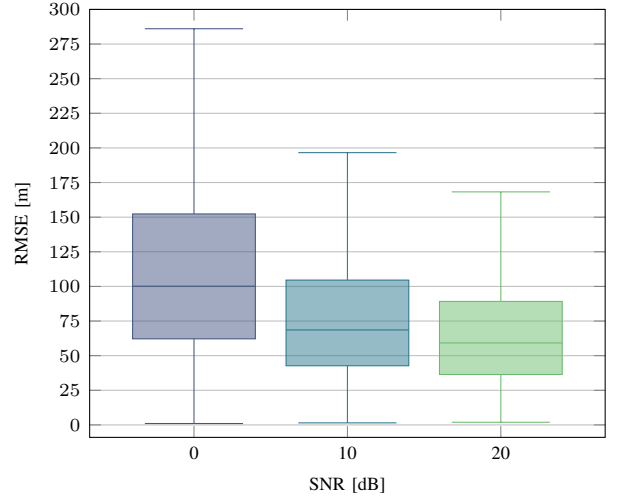
\begin{figure} 
    \centering
    \definecolor{darkslateblue}{RGB}{68,86,129}
\definecolor{mediumseagreen}{RGB}{107,187,110}
\definecolor{seagreen}{RGB}{46,130,127}
\definecolor{teal41120142}{RGB}{41,120,142}

\begin{tikzpicture}
  \begin{axis}
    [
    boxplot/draw direction=y,
    ylabel={RMSE [m]},
    xlabel={SNR [dB]},
    ymin=-9, ymax=300,
    ytick distance =25,
    cycle list={{darkslateblue},{teal41120142},{mediumseagreen}},
    xtick={1,2,3},
    ymajorgrids,
    xticklabels={0,10,20},
    xlabel style={font=\scriptsize}, ylabel style={font=\scriptsize}, ticklabel style={font=\scriptsize}
    ]
    \addplot+[
    fill,fill opacity=0.5, 
    boxplot prepared={
      median=100.15,
      upper quartile=152.33,
      lower quartile=62.15,
      upper whisker=286.05,
      lower whisker=1.01
    },
    ] coordinates {};

    \addplot+[
    fill,fill opacity=0.5, 
    boxplot prepared={
      median= 68.56,
      upper quartile=104.56,
      lower quartile=42.71,
      upper whisker=196.58,
      lower whisker=1.47
    },
    ] coordinates {};

    \addplot+[
    fill,fill opacity=0.5,  
    boxplot prepared={
      median=59.19,
      upper quartile=89.17,
      lower quartile=36.39,
      upper whisker=168.26,
      lower whisker=1.92
    },
    ] coordinates {};

    
  \end{axis}
\end{tikzpicture}
    \vspace{-1cm}\caption{Transmitter position errors distribution achieved by the LOC-NET, for different values of \ac{snr}.}
    \label{fig:boxplot_snr}
\end{figure}

Fig.~\ref{fig:boxplot_snr} shows (in box plots) the error distributions for the localization task as a function of the $\mathrm{SNR}$, for $\mathrm{SNR} = 0$, $10$, and $\SI{20}{\decibel}$. As expected, the average accuracy improves with increasing \ac{snr} furthermore, this leads to a decrease in the error confidence interval. 

To assess the localization accuracy of the standalone predictor block, we generated a trajectories dataset of source positions following the mobility model described in Section~\ref{subsec:bellhop}. In detail, we considered 
\begin{equation}\label{eq:trajSim}
    \bm{p}^\star(t) = \bm{p}(t) + \bm{w}_\mathrm{pos}\,,
\end{equation}
where $\bm{p}(t)$ is the true trajectory, simulated as in \eqref{eq:position}, and $\bm{w}_\mathrm{pos}$ is a Gaussian stationary noise with zero mean and \ac{std} values equal on each component $\sigma_\mathrm{pos}$, modeling the uncertainty introduced by the position estimator. In particular, we considered as possible \acp{std} $\sigma_\mathrm{pos} = 25$, $50$, and $100$~m.
Then both the \ac{rnn} and the Kalman filter were tested receiving as input trajectories $\bm{p}^\star(t)$.

\begin{figure} 
    \centering
    \definecolor{darkslateblue}{RGB}{68,86,129}
\definecolor{mediumseagreen}{RGB}{107,187,110}
\definecolor{seagreen}{RGB}{46,130,127}
\definecolor{teal41120142}{RGB}{41,120,142}

\begin{tikzpicture}
  \begin{axis}
    [
    boxplot/draw direction=y,
    ylabel={RMSE [m]},
    xlabel={Noise \ac{std} $\sigma_\mathrm{pos}$  [m]},
    ymin=-9, ymax=155,
    ytick distance =25,
    cycle list={{mediumseagreen},{teal41120142}},
    xtick={1,2,3},
    ymajorgrids,
    xticklabels={$25$, $50$, $100$},
    xlabel style={font=\scriptsize}, ylabel style={font=\scriptsize}, ticklabel style={font=\scriptsize},
    /pgfplots/boxplot/box extend=0.35
    ]
    \matrix[draw,  fill=white] at (.8,135) {  
        \node [shape=rectangle, draw=teal41120142 ,fill=teal41120142, opacity=.5, label=right:\scriptsize{Kalman}] {}; \\
        \node [shape=rectangle, draw=mediumseagreen ,fill=mediumseagreen, opacity=.5, label=right:\scriptsize{RNN}] {}; \\
    };
    \addplot+[
    fill,fill opacity=0.5, 
    boxplot prepared={
      median=21.76,
      upper quartile=31.19,
      lower quartile=13.94,
      upper whisker=56.97,
      lower whisker=0.11,
      draw position=0.8
    },
    ] coordinates {};

    \addplot+[
    fill,fill opacity=0.5, 
    boxplot prepared={
      median=21.82,
      upper quartile=28.89,
      lower quartile=15.09,
      upper whisker=49.50,
      lower whisker=0.38,
      draw position=1.2
    },
    ] coordinates {};

    \addplot+[
    fill,fill opacity=0.5,  
    boxplot prepared={
      median=31.55,
      upper quartile=46.58,
      lower quartile=19.998,
      upper whisker=86.43,
      lower whisker=0.23,
      draw position=1.8
    },
    ] coordinates {};

    \addplot+[
    fill,fill opacity=0.5, 
    boxplot prepared={
      median=28.92,
      upper quartile=42.49,
      lower quartile=18.36,
      upper whisker=78.67,
      lower whisker=0.36,
      draw position=2.2
    },
    ] coordinates {};

    \addplot+[
    fill,fill opacity=0.5, 
    boxplot prepared={
      median= 51.49,
      upper quartile=75.14,
      lower quartile=32.96,
      upper whisker=138.29,
      lower whisker=0.32,
      draw position=2.8
    },
    ] coordinates {};

    \addplot+[
    fill,fill opacity=0.5,  
    boxplot prepared={
      median=53.24,
      upper quartile=79.97,
      lower quartile=33.82,
      upper whisker=149.20,
      lower whisker=0.23,
      draw position=3.2
    },
    ] coordinates {};

    
  \end{axis}
\end{tikzpicture}
    \vspace{-1cm}\caption{Transmitter position errors of Kalman and \ac{rnn}, receiving as input noisy source trajectories, with position estimation noise $\sigma_\mathrm{pos}$.}
    \label{fig:boxplot_simpos}
    \vspace{-.2cm}
\end{figure}
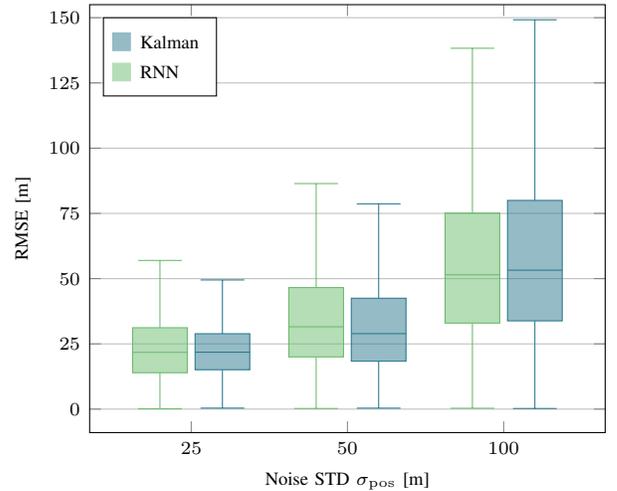

As shown in Fig.~\ref{fig:boxplot_simpos}, the proposed predictors achieve similar performance in mean at a high \ac{std} $\sigma_\mathrm{pos}$. 


\begin{figure} 
    \centering
    \definecolor{darkslateblue}{RGB}{68,86,129}
\definecolor{mediumseagreen}{RGB}{107,187,110}
\definecolor{seagreen}{RGB}{46,130,127}
\definecolor{teal41120142}{RGB}{41,120,142}

\begin{tikzpicture}
  \begin{axis}
    [
    boxplot/draw direction=y,
    ylabel={RMSE [m]},
    xlabel={Localization Model},
    ymin=-9, ymax=190,
    ytick distance =25,
    cycle list={{darkslateblue},{teal41120142},{mediumseagreen}},
    xtick={1,2,3},
    ymajorgrids,
    xticklabels={LOC-NET only, Kalman Filter, RNN},
    xlabel style={font=\scriptsize}, ylabel style={font=\scriptsize}, ticklabel style={font=\scriptsize}
    ]
    \addplot+[
    fill,fill opacity=0.5, 
    boxplot prepared={
      median=59.19,
      upper quartile=89.17,
      lower quartile=36.39,
      upper whisker=168.26,
      lower whisker=1.92
    },
    ] coordinates {};

    \addplot+[
    fill,fill opacity=0.5, 
    boxplot prepared={
      median= 49.28,
      upper quartile=75.48,
      lower quartile=31.64,
      upper whisker=141.08,
      lower whisker=0.67
    },
    ] coordinates {};

    \addplot+[
    fill,fill opacity=0.5,  
    boxplot prepared={
      median=45.93,
      upper quartile=71.72,
      lower quartile=27.96,
      upper whisker=137.20,
      lower whisker=0.36
    },
    ] coordinates {};

    
  \end{axis}
\end{tikzpicture}
    \vspace{-1cm}\caption{Transmitter position errors of LOC-NET, Kalman, and \gls{rnn} with ${\rm SNR} = \SI{20}{\decibel}$.}
    \label{fig:boxplot}
    \vspace{-.2cm}
\end{figure}
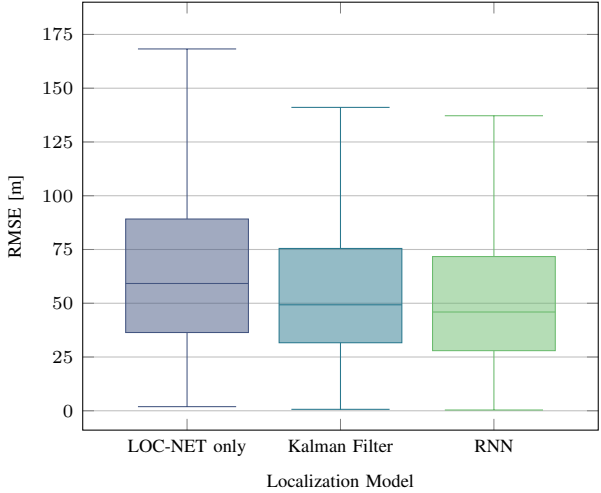

We now consider the localization performance at the output of the position predictor block obtained using the LOC-NET as the position estimator. Fig.~\ref{fig:boxplot} collects the results for $\mathrm{SNR} = \SI{20}{\decibel}$. Both the Kalman filter and the \ac{rnn}  improve the performance of the LOC-NET since they rely on the time sequence correlation to correct and smooth the localization errors induced by the LOC-NET.
In addition, it can be noticed that the \ac{rnn} achieves a slightly lower prediction error.

\begin{table}
    \caption{Estimated localization errors with ${\rm SNR} = \SI{20}{\decibel}$.}
    \centering
    \renewcommand{\arraystretch}{1.5} 
    \begin{tabular}{|c|c c c c|}
        \hline
        Loc. Model & x \acs{rmse} & y \acs{rmse}& x \acs{mape} & y \acs{mape}\\
         \hline
         LOC-NET only & $72.7$ m & $63.1$ m & $5.8\%$ & $4.9\%$ \\
         
         LOC-NET + Kalman & $63.4$ m & $54.7$ m & $4.9\%$ & $4.1\%$ \\
         
         LOC-NET + RNN &  $60.8$ m & $55.5$ m & $4.6\%$ & $4.1\%$ \\
         
        \hline
    \end{tabular}
    \label{tab:error_xy}
\end{table}

The numerical metrics reported in Table~\ref{tab:error_xy}, besides highlighting the great average results of the proposed estimator and predictors, confirm the fact that both \ac{rnn} and Kalman filter improve the LOC-NET estimates, achieving almost the same \ac{rmse} and \ac{mape} in estimating the source coordinates. 
Still, we remark that the Kalman filter is less general than the \gls{rnn} since it requires an a priori knowledge about the nature of the source motion. In this sense, it has an inherent advantage over the \ac{rnn}, which has instead no a priori knowledge about the source motion model plugged in. 

\begin{figure}
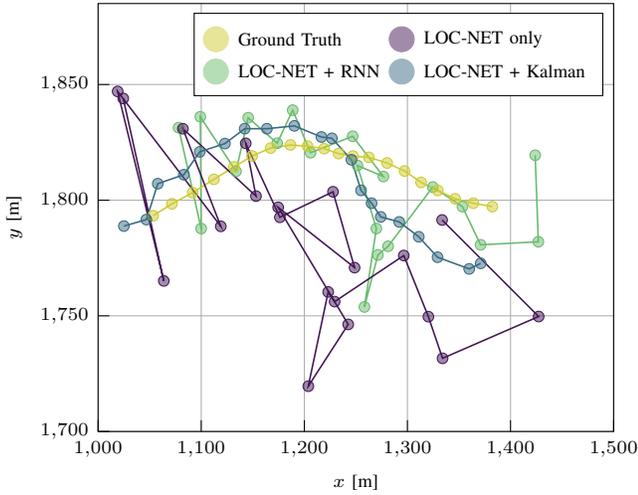

    \centering
    \include{plots/tx_path}
    \vspace{-1cm}\caption{Example of localized transmitter position using LOC-NET only, LOC-NET + Kalman, and LOC-NET + RNN, with ${\rm SNR} = \SI{20}{\decibel}$.}
    \label{fig:tx_path}
    \vspace{-.2cm}
\end{figure}

The effect of such knowledge is shown in  Fig.~\ref{fig:tx_path}, where the Kalman filter predictions are equally spaced and resemble the trajectory of the real source motion. This is because the Kalman filter forces the source motion model on the output, thus outputting a trajectory that is close to the ground truth since the real transmitter motion can be well approximated by a constant speed motion model. On the other hand, the source positions predicted by the \ac{rnn} do not have the same shape, as the \ac{rnn} only works using the LOC-NET (noisy) output. Still, this highlighted the effectiveness of the proposed approach that, even with an inherent disadvantage, it can achieve a slightly more accurate localization. 

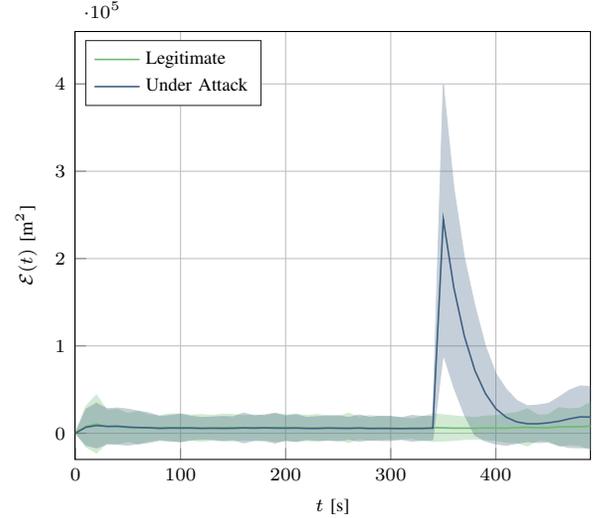
\begin{figure}
    \centering
\begin{tikzpicture}

\definecolor{darkgray176}{RGB}{176,176,176}
\definecolor{darkorange2551270}{RGB}{255,127,0}
\definecolor{steelblue31120180}{RGB}{31,120,180}

\definecolor{mediumseagreen}{RGB}{107,187,110}
\definecolor{darkslateblue}{RGB}{62,96,130}

\begin{axis}[
tick pos=left,
xlabel={$t$ [s]},
ytick distance = 1e5,
xmajorgrids,
xmin=-0.5, xmax=490,
ylabel={$\mathcal{E}(t)$ [m$^2$]},
ymajorgrids,
ymin=-3e4, ymax=4.6e5,
legend style={at={(0.02,0.98)}, anchor=north west, legend cell align=left, align=left, draw=white!15!black},
xlabel style={font=\scriptsize}, ylabel style={font=\scriptsize}, ticklabel style={font=\scriptsize},legend style={font=\scriptsize}
]
\path [draw=mediumseagreen, fill=mediumseagreen, opacity=0.3]
(axis cs:0,200.000000000001)
--(axis cs:0,199.999999999999)
--(axis cs:10,-15495.7027482308)
--(axis cs:20,-22654.9241369726)
--(axis cs:30,-11130.5416733225)
--(axis cs:40,-11998.6140258002)
--(axis cs:50,-9537.57869566997)
--(axis cs:60,-11943.3073513332)
--(axis cs:70,-8882.08092710164)
--(axis cs:80,-6955.28867697548)
--(axis cs:90,-9494.80200427422)
--(axis cs:100,-8834.39460759689)
--(axis cs:110,-7859.18564771699)
--(axis cs:120,-8975.81678957595)
--(axis cs:130,-8027.57162080535)
--(axis cs:140,-8944.72390352716)
--(axis cs:150,-8699.22862006535)
--(axis cs:160,-7615.45127462803)
--(axis cs:170,-7879.95010843358)
--(axis cs:180,-7536.87028019104)
--(axis cs:190,-9474.05577430571)
--(axis cs:200,-8328.82140969612)
--(axis cs:210,-9372.90338522652)
--(axis cs:220,-8316.34514714052)
--(axis cs:230,-9262.67598664711)
--(axis cs:240,-7922.73718038319)
--(axis cs:250,-7335.97305116943)
--(axis cs:260,-10622.4112478723)
--(axis cs:270,-6577.9989798236)
--(axis cs:280,-8310.26655651672)
--(axis cs:290,-7303.58449670967)
--(axis cs:300,-6868.16736180848)
--(axis cs:310,-6447.60011465195)
--(axis cs:320,-5955.81039348434)
--(axis cs:330,-8075.74580976914)
--(axis cs:340,-9295.87837622419)
--(axis cs:350,-8858.68372062111)
--(axis cs:360,-8200.11147733884)
--(axis cs:370,-7066.17582478916)
--(axis cs:380,-6508.74424394297)
--(axis cs:390,-7365.33109781574)
--(axis cs:400,-8197.42205148121)
--(axis cs:410,-10600.0880634871)
--(axis cs:420,-10808.9661128064)
--(axis cs:430,-14552.5866387885)
--(axis cs:440,-8343.90284513168)
--(axis cs:450,-8700.12425593795)
--(axis cs:460,-14593.3570886984)
--(axis cs:470,-13061.102019871)
--(axis cs:480,-13945.0787858979)
--(axis cs:490,-19001.7362759661)
--(axis cs:490,35512.6301275466)
--(axis cs:490,35512.6301275466)
--(axis cs:480,28737.3370825919)
--(axis cs:470,27960.7243472992)
--(axis cs:460,29317.9958506535)
--(axis cs:450,21257.6911222022)
--(axis cs:440,20915.7271866068)
--(axis cs:430,27989.2043887497)
--(axis cs:420,23717.0922776754)
--(axis cs:410,22290.8023876444)
--(axis cs:400,20132.0949131925)
--(axis cs:390,19588.2387070673)
--(axis cs:380,18002.6201965312)
--(axis cs:370,18807.6293271001)
--(axis cs:360,19854.2118500199)
--(axis cs:350,21111.4863012285)
--(axis cs:340,22159.3079442469)
--(axis cs:330,19092.3788561688)
--(axis cs:320,16399.7619233935)
--(axis cs:310,17835.2308540346)
--(axis cs:300,18814.4361658618)
--(axis cs:290,19035.1699900303)
--(axis cs:280,20369.7730226022)
--(axis cs:270,18516.6781926862)
--(axis cs:260,22887.4433441835)
--(axis cs:250,19218.4671192416)
--(axis cs:240,20116.646077987)
--(axis cs:230,21471.9296101159)
--(axis cs:220,20259.0498869274)
--(axis cs:210,22004.7835032254)
--(axis cs:200,20922.5910003879)
--(axis cs:190,22370.9588956201)
--(axis cs:180,19974.344983268)
--(axis cs:170,21105.8218388294)
--(axis cs:160,19975.9126865741)
--(axis cs:150,21526.5090740478)
--(axis cs:140,21778.690357733)
--(axis cs:130,20389.7938114177)
--(axis cs:120,21397.0266314682)
--(axis cs:110,20757.6528378995)
--(axis cs:100,22226.2152897157)
--(axis cs:90,22199.6138073265)
--(axis cs:80,19884.1783919949)
--(axis cs:70,21049.8432619947)
--(axis cs:60,24958.6210695514)
--(axis cs:50,22726.5517230345)
--(axis cs:40,27303.2332229611)
--(axis cs:30,26398.907609954)
--(axis cs:20,43837.6544784816)
--(axis cs:10,30959.620051949)
--(axis cs:0,200.000000000001)
--cycle;

\path [draw=darkslateblue, fill=darkslateblue, opacity=0.3]
(axis cs:0,200)
--(axis cs:0,200)
--(axis cs:10,-13594.3030052089)
--(axis cs:20,-16740.3021884302)
--(axis cs:30,-12350.8307976884)
--(axis cs:40,-12696.642717556)
--(axis cs:50,-13487.3619393843)
--(axis cs:60,-12077.2684600534)
--(axis cs:70,-10562.4391685673)
--(axis cs:80,-8208.96623227214)
--(axis cs:90,-8377.10163408065)
--(axis cs:100,-9979.7456538495)
--(axis cs:110,-7597.10851231969)
--(axis cs:120,-6883.58491041911)
--(axis cs:130,-7084.1852100023)
--(axis cs:140,-7976.47677798248)
--(axis cs:150,-7245.83502206102)
--(axis cs:160,-10977.1104217771)
--(axis cs:170,-8255.43717631905)
--(axis cs:180,-8645.06813801197)
--(axis cs:190,-11172.5158385857)
--(axis cs:200,-8091.3417999304)
--(axis cs:210,-9267.89713500284)
--(axis cs:220,-7147.8049285591)
--(axis cs:230,-7939.4634314133)
--(axis cs:240,-7902.16919065295)
--(axis cs:250,-8821.99637462469)
--(axis cs:260,-6871.93810840825)
--(axis cs:270,-9139.22214588075)
--(axis cs:280,-7490.85170763998)
--(axis cs:290,-8306.68083479166)
--(axis cs:300,-8369.52271269064)
--(axis cs:310,-7039.1738888573)
--(axis cs:320,-9027.63841708897)
--(axis cs:330,-8121.77337206891)
--(axis cs:340,-7350.00883882077)
--(axis cs:350,90637.61519976)
--(axis cs:360,52518.6482607142)
--(axis cs:370,21439.8973224593)
--(axis cs:380,-2187.34921641533)
--(axis cs:390,-8928.61806942458)
--(axis cs:400,-12279.5765777005)
--(axis cs:410,-13735.1581001455)
--(axis cs:420,-11564.8369144383)
--(axis cs:430,-9551.1332268457)
--(axis cs:440,-10616.8678492731)
--(axis cs:450,-10231.496549848)
--(axis cs:460,-13803.1249356985)
--(axis cs:470,-16307.685768964)
--(axis cs:480,-16610.4204986959)
--(axis cs:490,-16583.9787290075)
--(axis cs:490,53648.3476444611)
--(axis cs:490,53648.3476444611)
--(axis cs:480,54122.9024136842)
--(axis cs:470,49360.6515453492)
--(axis cs:460,41293.8191831252)
--(axis cs:450,34132.0068760061)
--(axis cs:440,32345.6988705932)
--(axis cs:430,31355.4056273874)
--(axis cs:420,37492.0865949487)
--(axis cs:410,50212.8085692116)
--(axis cs:400,68315.8667920209)
--(axis cs:390,100112.554204942)
--(axis cs:380,146128.782306243)
--(axis cs:370,201144.542032475)
--(axis cs:360,281004.787861286)
--(axis cs:350,398993.925375944)
--(axis cs:340,18998.5117789859)
--(axis cs:330,19371.9354376684)
--(axis cs:320,20356.4139455497)
--(axis cs:310,17759.9408381119)
--(axis cs:300,19312.4632788403)
--(axis cs:290,19462.6058704617)
--(axis cs:280,18261.9321758634)
--(axis cs:270,21005.92515192)
--(axis cs:260,18013.8056218699)
--(axis cs:250,20282.7125277708)
--(axis cs:240,19779.5540115252)
--(axis cs:230,18949.6554663251)
--(axis cs:220,18255.8844628267)
--(axis cs:210,21159.4751950312)
--(axis cs:200,19772.6117402483)
--(axis cs:190,22922.1756872638)
--(axis cs:180,20935.1185450201)
--(axis cs:170,19875.3732118268)
--(axis cs:160,23195.9419525427)
--(axis cs:150,18719.3868468348)
--(axis cs:140,19127.8657572669)
--(axis cs:130,18626.207338338)
--(axis cs:120,18033.3235033314)
--(axis cs:110,19472.154945916)
--(axis cs:100,21717.5394657634)
--(axis cs:90,20242.7019387581)
--(axis cs:80,19354.2893631562)
--(axis cs:70,23289.4164848442)
--(axis cs:60,25259.677220345)
--(axis cs:50,27618.3640899999)
--(axis cs:40,28902.436642759)
--(axis cs:30,28113.4713867162)
--(axis cs:20,34374.7095343583)
--(axis cs:10,27492.9308823676)
--(axis cs:0,200)
--cycle;

\addplot [semithick, mediumseagreen]
table {%
0 200
10 7731.95865185912
20 10591.3651707545
30 7634.18296831574
40 7652.30959858047
50 6594.48651368228
60 6507.65685910908
70 6083.88116744651
80 6464.4448575097
90 6352.40590152615
100 6695.91034105939
110 6449.23359509124
120 6210.60492094611
130 6181.11109530616
140 6416.98322710295
150 6413.64022699121
160 6180.23070597302
170 6612.9358651979
180 6218.73735153849
190 6448.4515606572
200 6296.88479534589
210 6315.94005899946
220 5971.35236989343
230 6104.62681173439
240 6096.95444880188
250 5941.24703403609
260 6132.5160481556
270 5969.33960643132
280 6029.75323304273
290 5865.7927466603
300 5973.13440202668
310 5693.81536969134
320 5221.97576495459
330 5508.31652319985
340 6431.71478401135
350 6126.40129030367
360 5827.05018634051
370 5870.72675115549
380 5746.93797629412
390 6111.4538046258
400 5967.33643085564
410 5845.35716207866
420 6454.06308243449
430 6718.30887498062
440 6285.91217073758
450 6278.78343313211
460 7362.31938097752
470 7449.81116371415
480 7396.12914834704
490 8255.44692579027
};
\addlegendentry{Legitimate};

\addplot [semithick, darkslateblue]
table {%
0 200
10 6949.31393857934
20 8817.20367296404
30 7881.32029451388
40 8102.89696260154
50 7065.50107530778
60 6591.20438014578
70 6363.48865813848
80 5572.66156544205
90 5932.8001523387
100 5868.89690595695
110 5937.52321679813
120 5574.86929645616
130 5771.01106416787
140 5575.69448964219
150 5736.77591238689
160 6109.41576538284
170 5809.96801775389
180 6145.02520350404
190 5874.82992433904
200 5840.63497015897
210 5945.78903001416
220 5554.03976713379
230 5505.09601745589
240 5938.69241043611
250 5730.35807657307
260 5570.93375673081
270 5933.3515030196
280 5385.5402341117
290 5577.96251783502
300 5471.47028307482
310 5360.38347462728
320 5664.38776423035
330 5625.08103279976
340 5824.25147008254
350 244815.770287852
360 166761.718061
370 111292.219677467
380 71970.716544914
390 45591.9680677585
400 28018.1451071602
410 18238.8252345331
420 12963.6248402552
430 10902.1362002708
440 10864.4155106601
450 11950.2551630791
460 13745.3471237134
470 16526.4828881926
480 18756.2409574941
490 18532.1844577268
};
\addlegendentry{Under Attack};

\end{axis}

\end{tikzpicture}
    \vspace{-1cm}\caption{Mean and $2\sigma$ confidence intervals of $\mathcal{E}(t)$, for the Kalman filter in the legitimate and under-attack (starting from $t=350$~s) scenarios, for ${\rm SNR} = \SI{20}{\decibel}$.}
    \label{fig:se_kalman}
    \vspace{-.2cm}
\end{figure}

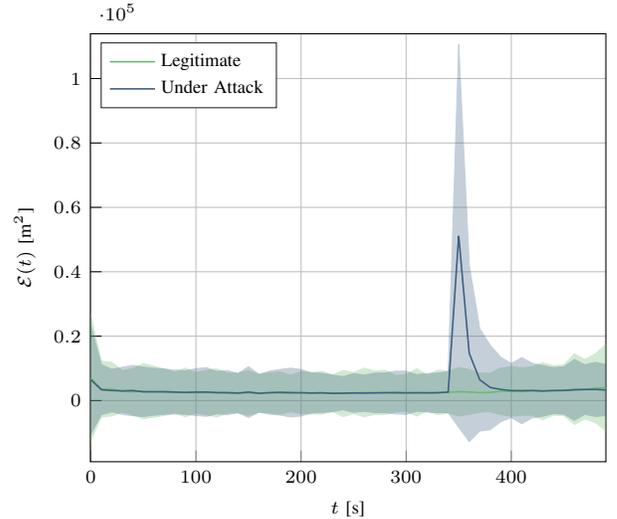
\begin{figure}
    \centering
\begin{tikzpicture}

\definecolor{darkgray176}{RGB}{176,176,176}
\definecolor{darkorange2551270}{RGB}{255,127,0}
\definecolor{steelblue31120180}{RGB}{31,120,180}

\definecolor{mediumseagreen}{RGB}{107,187,110}
\definecolor{darkslateblue}{RGB}{62,96,130}

\begin{axis}[
tick pos=left,
xlabel={$t$ [s]},
xmajorgrids,
xmin=-0.5, xmax=490,
ylabel={$\mathcal{E}(t)$ [m$^2$]},
ymajorgrids,
ymin=-1.9e4, ymax=113900.478654884,
ytick style={color=black},
legend style={at={(0.02,0.98)}, anchor=north west, legend cell align=left, align=left, draw=white!15!black},
xlabel style={font=\scriptsize}, ylabel style={font=\scriptsize}, ticklabel style={font=\scriptsize},legend style={font=\scriptsize}
]
\path [draw=mediumseagreen, fill=mediumseagreen, opacity=0.3]
(axis cs:0,26463.2444149857)
--(axis cs:0,-12384.7587308451)
--(axis cs:10,-5107.31123083463)
--(axis cs:20,-4807.25829005837)
--(axis cs:30,-3610.72457922696)
--(axis cs:40,-4469.84572047649)
--(axis cs:50,-5646.22229217874)
--(axis cs:60,-5178.10012301594)
--(axis cs:70,-4631.00677833645)
--(axis cs:80,-3789.09987187154)
--(axis cs:90,-5041.77065486712)
--(axis cs:100,-3832.60429874882)
--(axis cs:110,-4468.33327340489)
--(axis cs:120,-5070.0943236873)
--(axis cs:130,-3915.73580114186)
--(axis cs:140,-3874.66151191151)
--(axis cs:150,-4571.39863917504)
--(axis cs:160,-3712.23251992289)
--(axis cs:170,-3797.98435926833)
--(axis cs:180,-4272.93048775945)
--(axis cs:190,-3444.77886869629)
--(axis cs:200,-4048.95210690123)
--(axis cs:210,-4126.46047341752)
--(axis cs:220,-4245.53920948083)
--(axis cs:230,-3386.59038933493)
--(axis cs:240,-4974.83431291294)
--(axis cs:250,-4054.21614735902)
--(axis cs:260,-5024.11096115924)
--(axis cs:270,-4316.10900584729)
--(axis cs:280,-4334.98974038539)
--(axis cs:290,-3462.4004436897)
--(axis cs:300,-3294.94314602074)
--(axis cs:310,-4806.56805839063)
--(axis cs:320,-3268.23664822225)
--(axis cs:330,-4368.00708024677)
--(axis cs:340,-3935.59419764331)
--(axis cs:350,-4801.6840216146)
--(axis cs:360,-4290.74947433319)
--(axis cs:370,-3624.85560102359)
--(axis cs:380,-3460.31827428287)
--(axis cs:390,-4424.81879372334)
--(axis cs:400,-4860.13612438965)
--(axis cs:410,-4578.81278707727)
--(axis cs:420,-4545.70449944902)
--(axis cs:430,-5945.89536821455)
--(axis cs:440,-4530.14125616334)
--(axis cs:450,-4673.36806046935)
--(axis cs:460,-7416.23901123183)
--(axis cs:470,-5750.50986136818)
--(axis cs:480,-6805.21093664569)
--(axis cs:490,-9378.37767954281)
--(axis cs:490,17375.6251157518)
--(axis cs:490,17375.6251157518)
--(axis cs:480,14514.7541491591)
--(axis cs:470,12718.9771975626)
--(axis cs:460,14506.011691372)
--(axis cs:450,11048.8279297222)
--(axis cs:440,10806.9909051228)
--(axis cs:430,11738.1054614224)
--(axis cs:420,10476.2548066408)
--(axis cs:410,10149.1674662011)
--(axis cs:400,10527.7899510711)
--(axis cs:390,10141.3611447035)
--(axis cs:380,8505.75850612954)
--(axis cs:370,8523.77677127744)
--(axis cs:360,9597.57572910207)
--(axis cs:350,10272.1497921726)
--(axis cs:340,9180.88726387239)
--(axis cs:330,9438.40661653978)
--(axis cs:320,7894.6502290941)
--(axis cs:310,9898.52162381863)
--(axis cs:300,7976.13737224565)
--(axis cs:290,7847.95246038808)
--(axis cs:280,9415.66824141264)
--(axis cs:270,9169.22012685501)
--(axis cs:260,10064.4099059886)
--(axis cs:250,8681.84867282419)
--(axis cs:240,9318.52534505021)
--(axis cs:230,7955.51332065644)
--(axis cs:220,9042.96679149865)
--(axis cs:210,8684.21726502458)
--(axis cs:200,8977.2317907275)
--(axis cs:190,7856.79251040006)
--(axis cs:180,8910.34226844335)
--(axis cs:170,8616.99619771475)
--(axis cs:160,8235.38187308096)
--(axis cs:150,9262.40661067304)
--(axis cs:140,8659.3291994967)
--(axis cs:130,8470.02350228618)
--(axis cs:120,9869.06134959531)
--(axis cs:110,9388.8611365065)
--(axis cs:100,8663.34772468601)
--(axis cs:90,10072.4812082692)
--(axis cs:80,8828.59548080944)
--(axis cs:70,9813.34345496053)
--(axis cs:60,10650.5929716082)
--(axis cs:50,11443.0777754857)
--(axis cs:40,9915.61815731833)
--(axis cs:30,9257.64476856324)
--(axis cs:20,11885.5039197633)
--(axis cs:10,12210.3520058737)
--(axis cs:0,26463.2444149857)
--cycle;

\path [draw=darkslateblue, fill=darkslateblue, opacity=0.3]
(axis cs:0,23251.2695310256)
--(axis cs:0,-10340.1626009651)
--(axis cs:10,-4412.88751679423)
--(axis cs:20,-3490.42886082754)
--(axis cs:30,-4318.97516254937)
--(axis cs:40,-4709.98878961197)
--(axis cs:50,-4959.87748908818)
--(axis cs:60,-4567.22981766552)
--(axis cs:70,-4241.7581594167)
--(axis cs:80,-4189.42284188634)
--(axis cs:90,-4236.31324461998)
--(axis cs:100,-4634.79887633459)
--(axis cs:110,-4904.99690069098)
--(axis cs:120,-4125.59020284035)
--(axis cs:130,-4381.98856637665)
--(axis cs:140,-3704.47564237171)
--(axis cs:150,-5254.04250803806)
--(axis cs:160,-3542.59140677471)
--(axis cs:170,-4250.91767599521)
--(axis cs:180,-4496.60984611702)
--(axis cs:190,-5097.40869264985)
--(axis cs:200,-4784.78048272205)
--(axis cs:210,-3891.16476391527)
--(axis cs:220,-3663.95714376013)
--(axis cs:230,-3445.36146323795)
--(axis cs:240,-2862.85524966687)
--(axis cs:250,-3652.44514985463)
--(axis cs:260,-3328.65970661229)
--(axis cs:270,-3744.88571131837)
--(axis cs:280,-3761.62791346054)
--(axis cs:290,-4152.84392650821)
--(axis cs:300,-4302.26833466014)
--(axis cs:310,-3296.49662425347)
--(axis cs:320,-3569.25405093432)
--(axis cs:330,-3733.34123092259)
--(axis cs:340,-3741.04098985249)
--(axis cs:350,-8308.77386181048)
--(axis cs:360,-12729.1244568071)
--(axis cs:370,-9494.71191995404)
--(axis cs:380,-9051.83672829879)
--(axis cs:390,-6572.71067115791)
--(axis cs:400,-5307.72811359924)
--(axis cs:410,-7189.00896072013)
--(axis cs:420,-5361.48292491951)
--(axis cs:430,-4855.64836082621)
--(axis cs:440,-4075.51123416148)
--(axis cs:450,-4215.87180109818)
--(axis cs:460,-6147.16466568059)
--(axis cs:470,-4272.24679476215)
--(axis cs:480,-4943.7888227361)
--(axis cs:490,-4640.45971795386)
--(axis cs:490,11214.907521308)
--(axis cs:490,11214.907521308)
--(axis cs:480,11873.3582712765)
--(axis cs:470,11132.330357003)
--(axis cs:460,12787.506723741)
--(axis cs:450,10463.9282246023)
--(axis cs:440,10237.7474948204)
--(axis cs:430,10842.4473625769)
--(axis cs:420,11601.0330254019)
--(axis cs:410,13335.1710394947)
--(axis cs:400,11509.5460395715)
--(axis cs:390,13465.8308005489)
--(axis cs:380,17183.9911669438)
--(axis cs:370,22396.2131534024)
--(axis cs:360,42113.135117964)
--(axis cs:350,110569.843902288)
--(axis cs:340,9001.08170127598)
--(axis cs:330,8533.97967208313)
--(axis cs:320,8400.51646373335)
--(axis cs:310,8071.00651472253)
--(axis cs:300,9100.78729197316)
--(axis cs:290,9124.0463338305)
--(axis cs:280,8567.46264028227)
--(axis cs:270,8502.37943259574)
--(axis cs:260,7942.89544041515)
--(axis cs:250,8372.07002242263)
--(axis cs:240,7462.69059890865)
--(axis cs:230,7895.98352105019)
--(axis cs:220,8442.68454008305)
--(axis cs:210,8585.90110516863)
--(axis cs:200,9621.28864733818)
--(axis cs:190,10057.8407401184)
--(axis cs:180,9650.13076759322)
--(axis cs:170,9148.54124822039)
--(axis cs:160,8047.42650838837)
--(axis cs:150,10537.4621288828)
--(axis cs:140,8311.68892778047)
--(axis cs:130,9322.9904264298)
--(axis cs:120,9046.84899277256)
--(axis cs:110,10210.4413221788)
--(axis cs:100,9887.59279982978)
--(axis cs:90,9283.10942672548)
--(axis cs:80,9460.5249880622)
--(axis cs:70,9762.36363591735)
--(axis cs:60,10010.5662541515)
--(axis cs:50,10377.3527158747)
--(axis cs:40,10919.6055964796)
--(axis cs:30,10344.6696346189)
--(axis cs:20,9717.26690175221)
--(axis cs:10,11147.630646274)
--(axis cs:0,23251.2695310256)
--cycle;

\addplot [semithick, mediumseagreen]
table {%
0 7039.24284207027
10 3551.52038751954
20 3539.12281485245
30 2823.46009466814
40 2722.88621842092
50 2898.42774165348
60 2736.24642429615
70 2591.16833831204
80 2519.74780446895
90 2515.35527670106
100 2415.3717129686
110 2460.2639315508
120 2399.483512954
130 2277.14385057216
140 2392.33384379259
150 2345.503985749
160 2261.57467657904
170 2409.50591922321
180 2318.70589034195
190 2206.00682085189
200 2464.13984191314
210 2278.87839580353
220 2398.71379100891
230 2284.46146566075
240 2171.84551606863
250 2313.81626273258
260 2520.14947241467
270 2426.55556050386
280 2540.33925051362
290 2192.77600834919
300 2340.59711311245
310 2545.976782714
320 2313.20679043593
330 2535.1997681465
340 2622.64653311454
350 2735.23288527901
360 2653.41312738444
370 2449.46058512693
380 2522.72011592333
390 2858.27117549007
400 2833.82691334074
410 2785.17733956194
420 2965.27515359591
430 2896.10504660394
440 3138.42482447975
450 3187.72993462643
460 3544.88634007009
470 3484.23366809722
480 3854.77160625672
490 3998.62371810448
};
\addlegendentry{Legitimate};

\addplot [semithick, darkslateblue]
table {%
0 6455.55346503023
10 3367.37156473988
20 3113.41902046233
30 3012.84723603478
40 3104.80840343381
50 2708.73761339326
60 2721.668218243
70 2760.30273825033
80 2635.55107308793
90 2523.39809105275
100 2626.39696174759
110 2652.72221074392
120 2460.6293949661
130 2470.50093002658
140 2303.60664270438
150 2641.70981042235
160 2252.41755080683
170 2448.81178611259
180 2576.7604607381
190 2480.21602373425
200 2418.25408230807
210 2347.36817062668
220 2389.36369816146
230 2225.31102890612
240 2299.91767462089
250 2359.812436284
260 2307.11786690143
270 2378.74686063869
280 2402.91736341087
290 2485.60120366114
300 2399.25947865651
310 2387.25494523453
320 2415.63120639952
330 2400.31922058027
340 2630.02035571175
350 51130.5350202388
360 14692.0053305785
370 6450.75061672419
380 4066.07721932249
390 3446.56006469551
400 3100.90896298613
410 3073.08103938728
420 3119.77505024121
430 2993.39950087532
440 3081.11813032946
450 3124.02821175206
460 3320.17102903022
470 3430.04178112041
480 3464.78472427018
490 3287.22390167705
};
\addlegendentry{Under Attack};
\end{axis}

\end{tikzpicture}
    \vspace{-1cm}\caption{Mean and $2\sigma$ confidence intervals of $\mathcal{E}(t)$, for \ac{rnn} in the legitimate and under-attack (starting from $t=350$~s) scenarios, for ${\rm SNR} = \SI{20}{\decibel}$.}
    \label{fig:se_rnn}
    \vspace{-.2cm}
\end{figure}

\subsection{Authentication Results}
\label{sec:auth_results}
In this Section, we report the security performance of the proposed protocol, in terms of the squared error defined in~\eqref{eq:auth_metric}.

We run $1000$ simulations, half for legitimate and half for the under-attack scenario. Each simulation contains the trajectory sampled with period $T=\SI{10}{\second}$, using $N_{\rm rx} = 50$ receivers (unless differently specified). Concerning the under-attack scenario, we considered that for $t<\ell T$ the transmission is legitimate, while for $t \geq \ell T$ Bob receives signals from Eve only,  with $\ell = 35$.  

More in detail, we consider a scenario where Eve partially localizes Alice, places herself at $D = \SI{500}{\meter}$ from Alice, and tries to replicate her motion pattern. 
Specifically, when Eve starts transmitting,
\begin{itemize}
    \item her position is drawn at random at a distance of $\SI{500}{\meter}$ from Alice's last position;
    \item her initial speed is $v(0)=\SI{1}{\meter/\second}$;
    \item the angle of her initial motion direction with respect to the positive $x$-axis is uniformly sampled in the interval $[\phi - \frac{\pi}{4},\phi + \frac{\pi}{4}]$, where $\phi$ is Alice's motion direction angle;
    \item her next positions are defined by the motion model in~\eqref{eq:position}.
\end{itemize}

To show the distribution of $\mathcal{E}(t)$, for each time $t$  Fig.s~\ref{fig:se_kalman} and~\ref{fig:se_rnn} show an interval having as endpoints $\mu - 2\sigma$ and $\mu + 2 \sigma$, with $\mu$ being the average and $\sigma$ being the standard deviation of the squared error distribution. Assuming the data to be Gaussian, the interval contains approximately $95\%$ of the data. The two figures are obtained using the Kalman filter and the \ac{rnn}, respectively, for position prediction; both figures are obtained with ${\rm SNR} =20$~dB. 

As expected, in both Fig.s~\ref{fig:se_kalman} and~\ref{fig:se_rnn}, the legitimate $\mathcal{E}(t)$ remains almost constant over time, while as the attack starts at $t=\SI{350}{\second}$ we see a peak. Thus, both techniques can detect the attack and are suitable for authentication purposes.
However, there are relevant differences between the two position predictor implementations. 
In particular, due to the previously-mentioned sensitivity to noise on input positions, differently from the Kalman filter the $2\sigma$ intervals in the legitimate and under-attack case are not separable when using the \ac{rnn}. 

\begin{figure}
        \centering
        \include{plots/DET}    
        \vspace{-1cm}\caption{\Ac{det} curves for the proposed Kalman and \ac{rnn} based position predictors, for ${\rm SNR} = 10$ and $\SI{20}{\decibel}$.}
        \label{fig:DET}
        \vspace{-.2cm}
\end{figure}   
\begin{figure}
    \centering
    \input{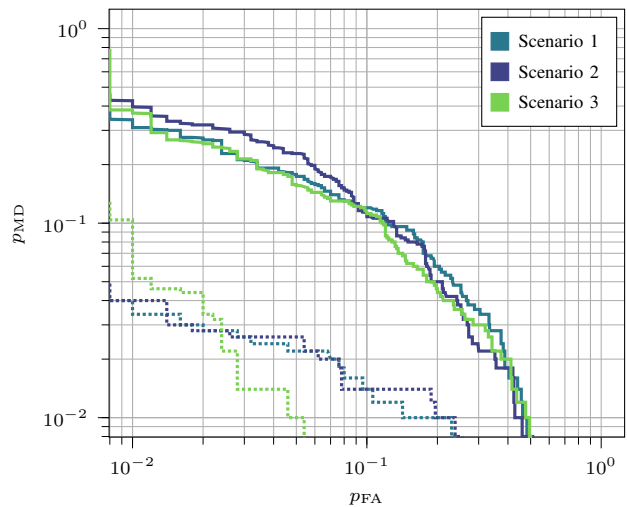}
    \caption{\ac{det} curves for the proposed \ac{rnn} and Kalman (dotted) based position predictors for ${\rm SNR}=\SI{20}{\decibel}$. In scenarios 1 and 2 environmental variability is considered for bathymetries 1 and 2, while scenario 3 is a reference with no environmental variability for bathymetry 1.}
    \label{fig:det_env}
    \vspace{-.2cm}
\end{figure}

To further investigate how such non-separability between the legitimate and under-attack $\mathcal{E}(t)$ affects the two predictors' ability to distinguish between Alice and Eve,  Fig.~\ref{fig:DET} shows the \ac{det} curves, i.e.,  the MD probability $p_\mathrm{MD}$ ~\eqref{eq:pmd} as a function of the FA probability $p_\mathrm{FA}$ ~\eqref{eq:pfa} for ${\rm SNR} = 10$ and $\SI{20}{\decibel}$. The results are obtained using $N_\mathrm{a}=10$, meaning that the decision $\hat{\mathcal{H}}(t)$ is taken only after $t > \SI{100}{\second}$. We also tested other values of $N_\mathrm{a}$, concluding that this parameter has a negligible impact on the performance. As expected, for both implementations better results are achieved at a higher \ac{snr}.
Still, the Kalman filter outperforms the \ac{rnn} and achieves $p_\mathrm{FA}$ and $p_\mathrm{MD}$ values of $0.024$ while the latter cannot reach values lower than $0.1$ for both probabilities.
These results are consistent with the previously presented $\mathcal{E}(t)$ for the Kalman filter and the \ac{rnn}, as the regularization imposed by the Kalman filter improves the detection of jumps on the position estimation. On the other hand, the  \ac{rnn} is trained to be robust to these inconsistencies, making room for Eve to lead a successful attack.

To address the variability of environmental conditions, we modify the Thorp absorption coefficient as a function of various parameters. More in detail, water's salinity, temperature, and pH take values respectively in $[30{\rm ppt},35{\rm ppt}]$, $[12^\circ{\rm C},24^\circ{\rm C}]$, and $[6,9]$. By changing these parameters for each testing trajectory, we obtained the authentication performance shown in Fig.~\ref{fig:det_env} for both available bathymetries. Note that the network has been trained on a single set of parameters, thus we are assessing its robustness against variations of the parameters with respect to the training phase. As expected, the environmental variability affects the LOC-NET accuracy and, consequently, the authentication performance. However, our model is robust even to high variations of the considered parameters.

\begin{figure}
    \centering
    \input{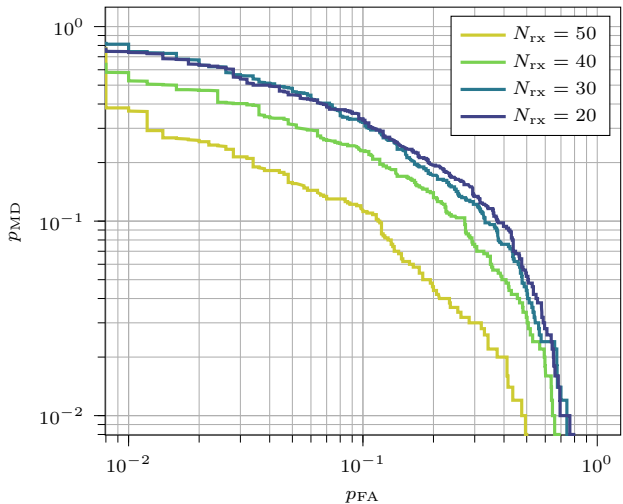}
    \caption{\ac{det} curves for the \ac{rnn} based position predictor with varying number of available receivers and ${\rm SNR}=\SI{20}{\decibel}$.}
    \label{fig:det_nrx_rnn}
\end{figure}

\begin{figure}
    \centering
    \input{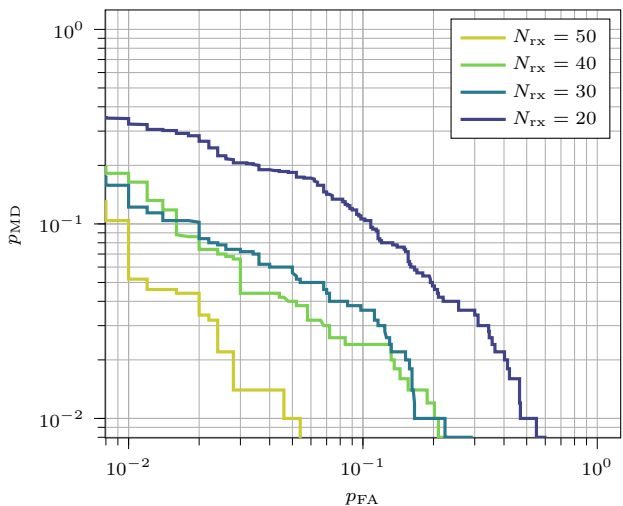}
    \caption{\ac{det} curves for the Kalman-based position predictor with varying the number of available receivers and ${\rm SNR}=\SI{20}{\decibel}$.}
    \label{fig:det_nrx_k}
    \vspace{-.2cm}
\end{figure}

\ac{uwan} can vary greatly in size and number of nodes, thus we tested the proposed authentication method scalability by varying the number of available receivers. Fig.s~\ref{fig:det_nrx_rnn} and~\ref{fig:det_nrx_k} show that our model is affected by the reduction of the available receivers, nonetheless achieving acceptable authentication performances even with as few as $N_{\rm rx}=20$.

As for the localization task, we tested the authentication performances removing the dependency on the LOC-NET estimates by generating transmitter positions as in \eqref{eq:trajSim} with $\sigma_\mathrm{pos}=25$, $50$, and \SI{100}{\meter}. 
To do so, we generated $1000$ source trajectories, half for the legitimate and half for the under-attack scenario, for each \ac{std} value.
The generated trajectories are then fed to the Kalman filter and the trained \ac{rnn} to extract $\mathcal{E}(t)$.

\begin{figure}
    \centering
    \input{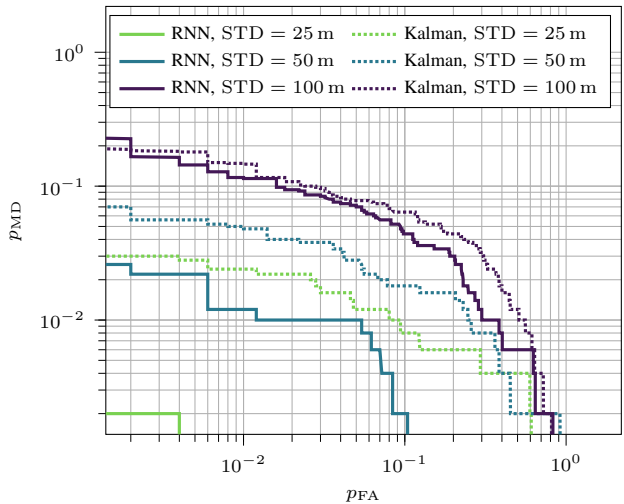}
    \caption{\ac{det} simulated positions with additive Gaussian noise.}
    \label{fig:det_simpos}
    \vspace{-.2cm}
\end{figure}

In Fig.~\ref{fig:det_simpos}, are plotted the \ac{det} curves representing the $p_\mathrm{MD}$ and the $p_\mathrm{FA}$ for the Kalman filter, the \ac{rnn} and for different values of the added Gaussian noise \ac{std}. When $\sigma_\mathrm{pos}=\SI{100}{\meter}$ the performances of the two models are almost the same. However, decreasing the noise \ac{std}, the \ac{rnn} outperforms the Kalman filter. This is because the noise samples added to the simulated positions, are uncorrelated between them while this is not the case for the LOC-NET errors. In fact, LOC-NET errors depend on the source location within the transmitter area therefore, the errors during one trajectory are correlated.

This error correlation favors the Kalman filter which can approximate the noisy trajectory with a constant speed motion model. On the other hand, having uncorrelated additive noise makes it more difficult to approximate the noisy trajectory with a constant speed motion model resulting in a less precise authentication metric $\mathcal{E}(t)$, which finally leads to higher $p_\mathrm{MD}$ and $p_\mathrm{FA}$. The \ac{rnn} instead, not having any prior assumption on the motion model, performs very well, especially with lower noise \ac{std}.
In conclusion, when the sea is calm and the source moves at a constant speed (at least with a good approximation), the Kalman filter is the best solution. On the other hand, in rough seas or when the source undergoes sudden acceleration, the \ac{rnn} should be preferred. In addition, while the Kalman filter typically outperforms the \ac{rnn}, this is more robust at lower \ac{snr}, with fewer receivers, and under higher environmental variability. Thus, these conditions should also be considered together with the source type of motion when choosing the position predictor.

 
\section{Conclusion}\label{sec:concl}

We have proposed a context-based authentication for \acp{uwan}, where an array of receivers aims at authenticating a moving transmitter relying on the physical properties of the underwater acoustic channel.
The proposed approach is divided into two steps. First, we estimate the source coordinates by using the LOC-NET, a novel \ac{cnn} developed for this task, which leverages the \ac{scm} obtained by combining the receivers' channel measurements.
Next, we use a position predictor to track the evolution of the source position. As a predictor, we have considered both a Kalman filter and a \ac{rnn}.

Finally, the security check relies on the comparison between the measured position with the predicted one, thus associating high position prediction errors with fake signals. 
The performance of the proposed scheme has been evaluated using a dataset collected from a Bellhop simulator. Results show that indeed the proposed scheme can distinguish between legitimate and fake signals while the transmitter is moving according to a realistic motion model. 

Specifically, the Kalman filter achieves great results as long as the source motion can be approximated with a constant speed motion model. 
However, in scenarios where the source speed cannot be approximated as constant, the \ac{rnn} would be a more appropriate predictor. 

Future works include the implementation of a dynamic threshold for detection and a sea trial to evaluate the performance of the proposed scheme in a real-world context. This may include, for instance, marine life and geological features that cannot be modeled via simulation.


\bibliography{biblio.bib}

\begin{thebibliography}{10}
\providecommand{\url}[1]{#1}
\csname url@samestyle\endcsname
\providecommand{\newblock}{\relax}
\providecommand{\bibinfo}[2]{#2}
\providecommand{\BIBentrySTDinterwordspacing}{\spaceskip=0pt\relax}
\providecommand{\BIBentryALTinterwordstretchfactor}{4}
\providecommand{\BIBentryALTinterwordspacing}{\spaceskip=\fontdimen2\font plus
\BIBentryALTinterwordstretchfactor\fontdimen3\font minus
  \fontdimen4\font\relax}
\providecommand{\BIBforeignlanguage}[2]{{%
\expandafter\ifx\csname l@#1\endcsname\relax
\typeout{** WARNING: IEEEtran.bst: No hyphenation pattern has been}%
\typeout{** loaded for the language `#1'. Using the pattern for}%
\typeout{** the default language instead.}%
\else
\language=\csname l@#1\endcsname
\fi
#2}}
\providecommand{\BIBdecl}{\relax}
\BIBdecl

\bibitem{lal2016secure}
C.~Lal, R.~Petroccia, M.~Conti, and J.~Alves, ``Secure underwater acoustic
  networks: Current and future research directions,'' in \emph{Proc. 3rd IEEE
  Underwater Commun. and Networking Conf. (UComms)}, 2016, pp. 1--5.

\bibitem{yang2019challenges}
G.~Yang, L.~Dai, G.~Si, S.~Wang, and S.~Wang, ``Challenges and security issues
  in underwater wireless sensor networks,'' in \emph{Proc. of the Int. Conf. on
  Identification, Inf. and Knowl. in the Internet of Things (IIKI)}, vol.
  147.\hskip 1em plus 0.5em minus 0.4em\relax Elsevier, 2019, pp. 210--216.

\bibitem{Waquas23Security}
W.~Aman, S.~Al-Kuwari, M.~Muzzammil, M.~M.~U. Rahman, and A.~Kumar, ``Security
  of underwater and air–water wireless communication: State-of-the-art,
  challenges and outlook,'' \emph{Ad Hoc Networks}, vol. 142, pp. 103--114,
  Apr. 2023.

\bibitem{diamant2018cooperative}
R.~Diamant, P.~Casari, and S.~Tomasin, ``Cooperative authentication in
  underwater acoustic sensor networks,'' \emph{IEEE Trans. on Wireless
  Commun.}, vol.~18, no.~2, pp. 954--968, Dec. 2018.

\bibitem{khalid_auth_aoa_mahala_oceans_2020}
M.~Khalid, R.~Zhao, and N.~Ahmed, ``Physical layer authentication in
  line-of-sight underwater acoustic sensor networks,'' in \emph{Proc. of
  OCEANS}.\hskip 1em plus 0.5em minus 0.4em\relax Singapore: IEEE, 2020, pp.
  1--5.

\bibitem{zhao22physical}
R.~Zhao, M.~Khalid, O.~A. Dobre, and X.~Wang, ``Physical layer node
  authentication in underwater acoustic sensor networks using time-reversal,''
  \emph{IEEE Sens. J.}, vol.~22, no.~4, pp. 3796--3809, Jan. 2022.

\bibitem{Zhao2023Physical}
R.~Zhao, T.~Shi, C.~Liu, X.~Shen, and O.~A. Dobre, ``Physical layer
  authentication without adversary training data in resource-constrained
  underwater acoustic networks,'' \emph{IEEE Sens. J.}, vol.~23, no.~22, pp.
  28\,270--28\,281, Nov. 2023.

\bibitem{bragagnolo2021authentication}
L.~Bragagnolo, F.~Ardizzon, N.~Laurenti, P.~Casari, R.~Diamant, and S.~Tomasin,
  ``Authentication of underwater acoustic transmissions via machine learning
  techniques,'' in \emph{Proc. IEEE Int. Conf. on Microw., Antennas, Commun.
  and Electron. Syst. (COMCAS)}, 2021, pp. 255--260.

\bibitem{du2022ltrust}
J.~Du, G.~Han, C.~Lin, and M.~Mart{\'\i}nez-Garc{\'\i}a, ``{LT}rust: An
  adaptive trust model based on {LSTM} for underwater acoustic sensor
  networks,'' \emph{IEEE Trans. on Wireless Commun.}, vol.~21, no.~9, pp.
  7314--7328, Sept. 2022.

\bibitem{zhang2023recommendation}
M.~Zhang, R.~Feng, H.~Zhang, and Y.~Su, ``A recommendation management defense
  mechanism based on trust model in underwater acoustic sensor networks,''
  \emph{Future Gener. Comput. Syst.}, vol. 145, pp. 466--477, Aug. 2023.

\bibitem{casari2022physical}
P.~Casari, F.~Ardizzon, and S.~Tomasin, ``Physical layer authentication in
  underwater acoustic networks with mobile devices,'' in \emph{Proc. of the
  16th Int. Conf. on Underwater Networks \& Systems (WUWNet)}, 2022.

\bibitem{Ardizzon2024RNN}
F.~Ardizzon, P.~Casari, and S.~Tomasin, ``A {RNN}-based approach to physical
  layer authentication in underwater acoustic networks with mobile devices,''
  \emph{Comput. Netw.}, vol. 243, p. 110311, Apr. 2024.

\bibitem{Aman2024novel}
W.~Aman, S.~Al-Kuwari, and M.~Qaraqe, ``A novel physical layer authentication
  mechanism for static and mobile {3D} underwater acoustic communication
  networks,'' \emph{Physical Communication}, vol.~66, p. 102430, Oct. 2024.

\bibitem{sazontov2015matched}
A.~Sazontov and A.~Malekhanov, ``Matched field signal processing in underwater
  sound channels,'' \emph{Acoustical Physics}, vol.~61, pp. 213--230, Mar.
  2015.

\bibitem{bellhop}
M.~B. Porter, ``Bellhop {G}aussian beam/finite element beam code,''
  \url{http://oalib.hlsresearch.com/Rays/}, (Last access August 2024).

\bibitem{niu2017source}
H.~Niu, E.~Reeves, and P.~Gerstoft, ``Source localization in an ocean waveguide
  using supervised machine learning,'' \emph{The Jour. of the Acoustical
  Society of America}, vol. 142, no.~3, pp. 1176--1188, Sept. 2017.

\bibitem{wang2018underwater}
Y.~Wang and H.~Peng, ``Underwater acoustic source localization using
  generalized regression neural network,'' \emph{The Jour. of the Acoustical
  Society of America}, vol. 143, no.~4, pp. 2321--2331, Apr. 2018.

\bibitem{huang2018source}
Z.~Huang, J.~Xu, Z.~Gong, H.~Wang, and Y.~Yan, ``Source localization using deep
  neural networks in a shallow water environment,'' \emph{The Jour. of the
  Acoustical Society of America}, vol. 143, no.~5, pp. 2922--2932, May 2018.

\bibitem{liu2020source}
W.~Liu, Y.~Yang, M.~Xu, L.~L{\"u}, Z.~Liu, and Y.~Shi, ``Source localization in
  the deep ocean using a convolutional neural network,'' \emph{The Jour. of the
  Acoustical Society of America}, vol. 147, no.~4, pp. 314--319, Apr. 2020.

\bibitem{liu2020multi}
Y.~Liu, H.~Niu, and Z.~Li, ``A multi-task learning convolutional neural network
  for source localization in deep ocean,'' \emph{The Jour. of the Acoustical
  Society of America}, vol. 148, no.~2, pp. 873--883, Aug. 2020.

\bibitem{9430737}
P.-F. Lv, B.~He, and J.~Guo, ``Position correction model based on gated hybrid
  {RNN} for {AUV} navigation,'' \emph{IEEE Trans. on Vehicular Technology},
  vol.~70, no.~6, pp. 5648--5657, May 2021.

\bibitem{qin2020underwater}
D.~Qin, J.~Tang, and Z.~Yan, ``Underwater acoustic source localization using
  {LSTM} neural network,'' in \emph{Proc. of the 39th Chinese Control
  Conference (CCC)}.\hskip 1em plus 0.5em minus 0.4em\relax IEEE, 2020, pp.
  7452--7457.

\bibitem{zhu2022time}
X.~Zhu, H.~Dong, P.~S. Rossi, and M.~Landr{\o}, ``Time-frequency fused
  underwater acoustic source localization based on contrastive predictive
  coding,'' \emph{IEEE Sensors Journal}, vol.~22, no.~13, pp. 13\,299--13\,308,
  June 2022.

\bibitem{Chollet_2017_CVPR}
F.~Chollet, ``Xception: Deep learning with depthwise separable convolutions,''
  in \emph{Proc. of the Conf. on Comput. Vis. and Pattern Recognit. (CVPR)},
  July 2017.

\bibitem{ioffe2015batch}
S.~Ioffe and C.~Szegedy, ``Batch normalization: Accelerating deep network
  training by reducing internal covariate shift,'' in \emph{Proc. Int. Conf. on
  machine learning}, 2015, pp. 448--456.

\bibitem{hao2020role}
W.~Hao, W.~Yizhou, L.~Yaqin, and S.~Zhili, ``The role of activation function in
  {CNN},'' in \emph{Proc. 2nd IEEE Int. Conf. on Inf. Technol. and Comput.
  Appl. (ITCA)}, 2020, pp. 429--432.

\bibitem{srivastava2014dropout}
N.~Srivastava, G.~Hinton, A.~Krizhevsky, I.~Sutskever, and R.~Salakhutdinov,
  ``Dropout: a simple way to prevent neural networks from overfitting,''
  \emph{Jour. of Machine Learning Research}, vol.~15, no.~1, pp. 1929--1958,
  June 2014.

\bibitem{welch1995introduction}
G.~Welch and G.~Bishop, ``An introduction to the {K}alman filter,''
  \emph{Technical report TR95-041}, 1995.

\bibitem{Kay:1993}
S.~Kay, \emph{Fundamentals of Statistical Signal Processing: Estimation
  Theory}.\hskip 1em plus 0.5em minus 0.4em\relax Englewood Cliffs, NJ:
  Prentice-Hall, 1993.

\bibitem{sak2014long}
H.~Sak, A.~Senior, and F.~Beaufays, ``Long short-term memory based recurrent
  neural network architectures for large vocabulary speech recognition,''
  \emph{arXiv}, Feb. 2014.

\bibitem{weerakody2021review}
P.~B. Weerakody, K.~W. Wong, G.~Wang, and W.~Ela, ``A review of irregular time
  series data handling with gated recurrent neural networks,''
  \emph{Neurocomputing}, vol. 441, pp. 161--178, June 2021.

\bibitem{porter1994finite}
M.~B. Porter and Y.-C. Liu, ``Finite-element ray tracing,'' \emph{Theoretical
  and computational acoustics}, vol.~2, pp. 947--956, Oct. 1994.

\end{thebibliography}
\bibliographystyle{IEEEtran}

\begin{IEEEbiography}[{\includegraphics[width=1in,height=1.25in,clip]{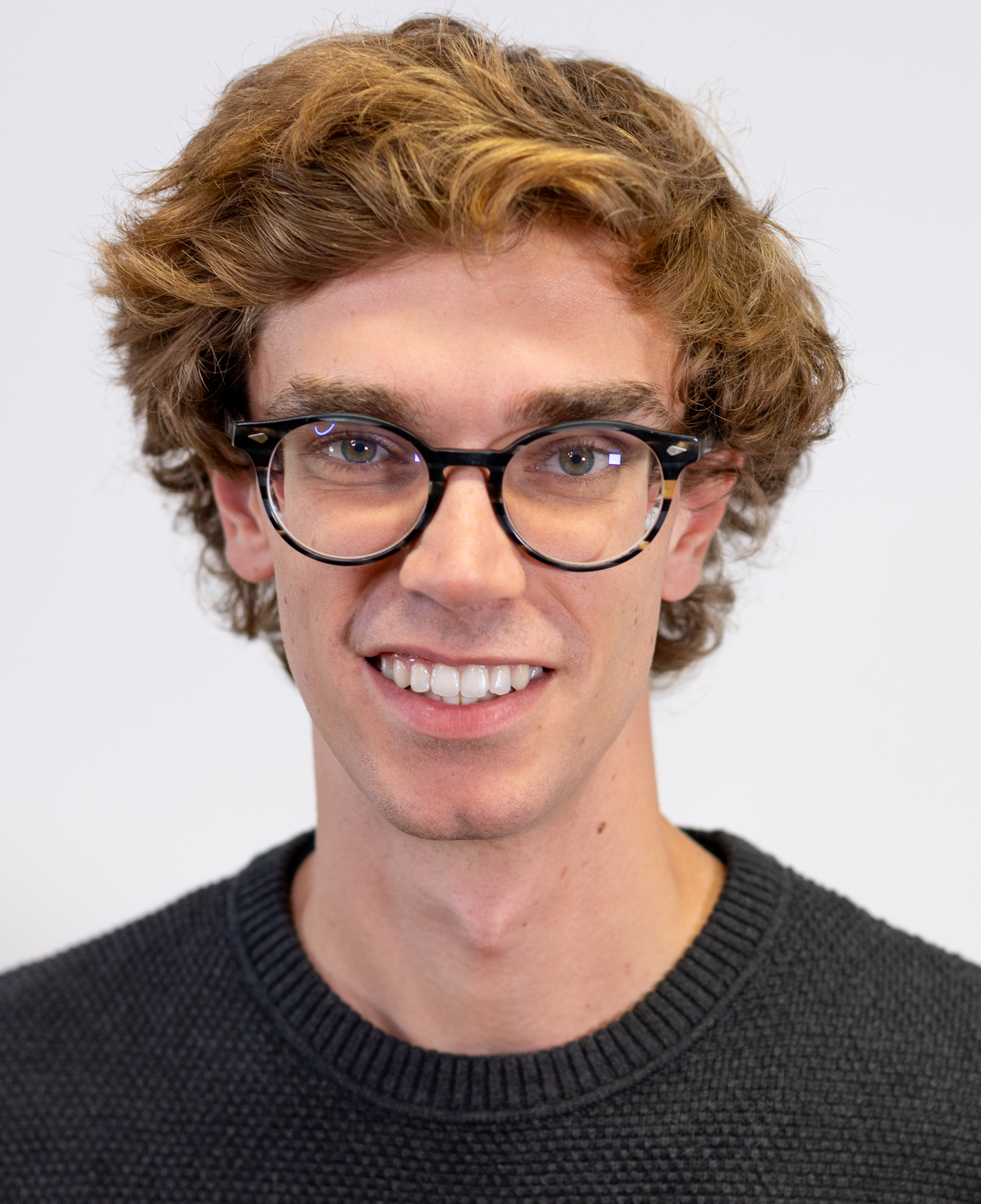}}]%
{Gianmaria Ventura}{\space}(Student Member, IEEE) 
received the B.Sc. degree in Information Engineering and the M.Sc. degree in Telecommunications Engineering from the University of Padova, Italy, in 2021 and 2023 respectively, where he is currently pursuing the Ph.D. degree in Information Engineering with the Department of Information Engineering, under the supervision of Prof. M. Rossi. His current research interests include authentication techniques for physical layer security and wireless joint communications and sensing.  
\vspace{-1cm}
\end{IEEEbiography}
\begin{IEEEbiography}[{\includegraphics[width=1in,height=1.25in,clip,keepaspectratio]{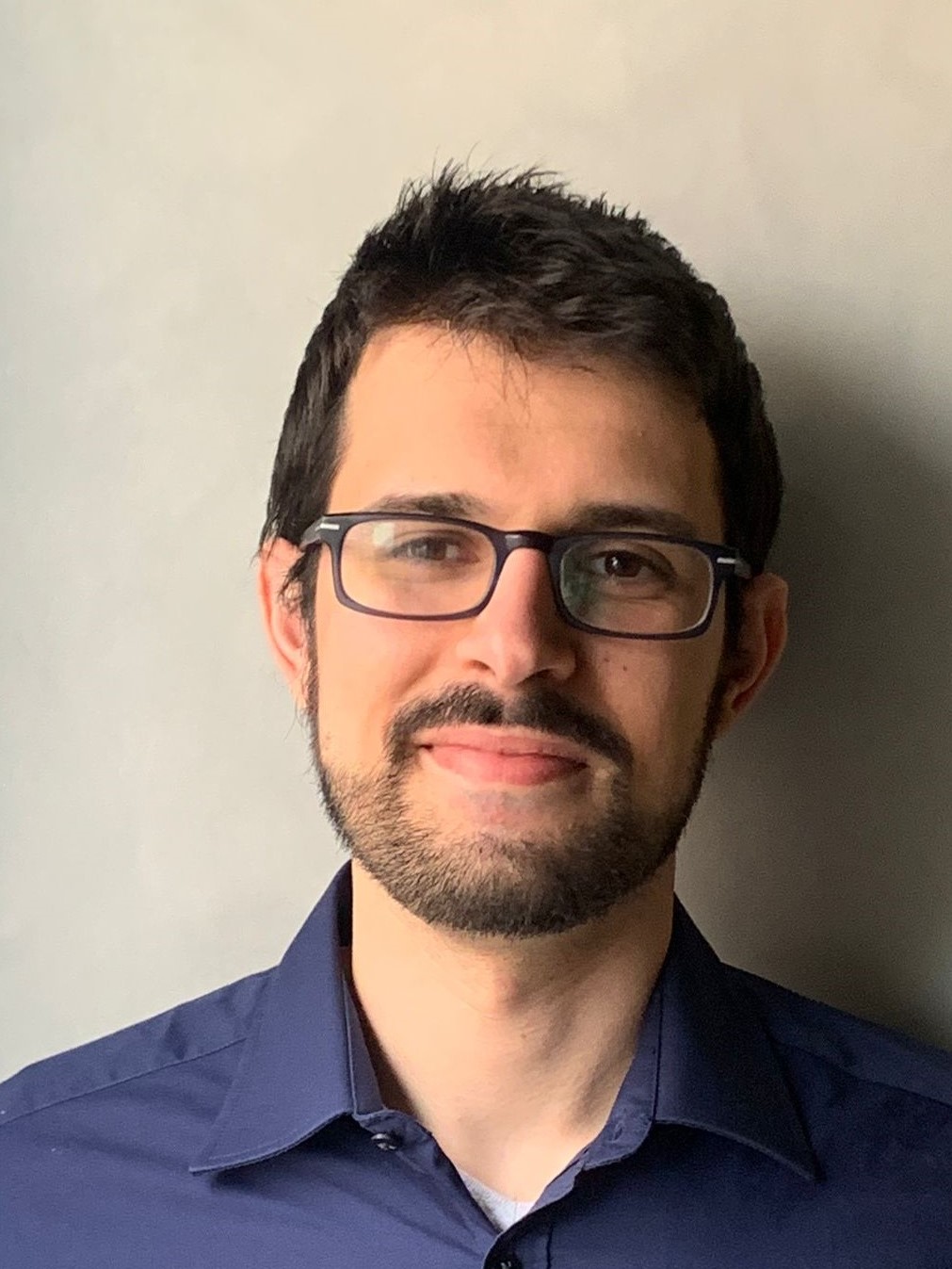}}]%
{Francesco Ardizzon}{\space}(Member, IEEE) received the B.Sc. degree in 2016, the M.Sc. degree in 2019, and the Ph.D. degree in Information Engineering in 2023 from the University of Padova, Italy. In 2022 he was a visiting scientist at the ESA European Space Research and Technology Centre. He is currently an Assistant Professor at the University of Padova. His current research interests include authentication for global navigation satellite systems, physical layer security, and underwater acoustic communications.
\vspace{-1cm}
\end{IEEEbiography}
\begin{IEEEbiography}
[{\includegraphics[width=1in,height=1.25in,clip,keepaspectratio]{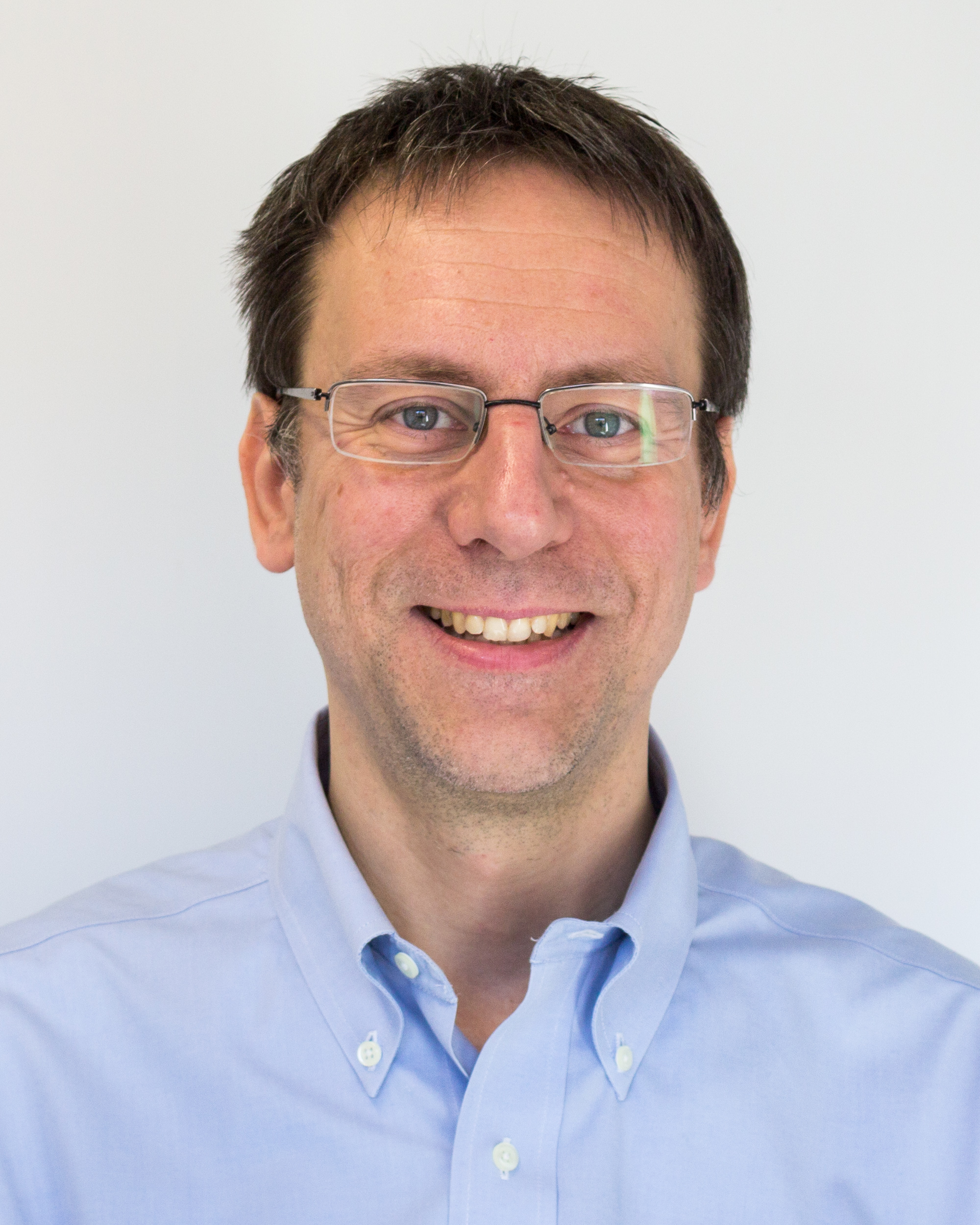}}]%
{Stefano Tomasin}{\space}(Senior Member, IEEE) received a Ph.D. degree from the University of Padova, Italy, in 2003. He joined the University of Padova where he is now Full Professor (since 2022). He was visiting faculty at Qualcomm, San Diego (CA) in 2004, the Polytechnic University in Brooklyn (NY) in 2007, and the Mathematical and Algorithmic Sciences Laboratory of Huawei in Paris (France) in 2015. His current research interests include physical layer security, security of global navigation satellitesystems, signal processing for wireless communications, synchronization, and scheduling of communication resources. He has been a senior member of IEEE since 2011 (member since 1999) and a member of EURASIP since 2011. He is a Deputy Editor-in-Chief of the IEEE Transactions on Information Forensics and Security since January 2023.

\end{IEEEbiography}




\end{document}